\documentclass[10pt,journal,compsoc]{IEEEtran}
\ifCLASSOPTIONcompsoc
  \usepackage[nocompress]{cite}
\else
  \usepackage{cite}
\fi
\usepackage{color}

\newcommand{\wgcb}[1]{\textcolor{black}{#1}}
\newcommand{\wgcr}[1]{\textcolor{black}{#1}}

\ifCLASSINFOpdf
   \usepackage[pdftex]{graphicx}
\else
\fi
\usepackage{amsmath}
\usepackage{array}
\begin{document}
%
\title{Light Field Reconstruction Using Convolutional Network on EPI and Extended Applications}
%
%
%
%

\author{Gaochang~Wu,~
        Yebin~Liu,~\IEEEmembership{Member,~IEEE,}
				Lu~Fang,~\IEEEmembership{Senior Member,~IEEE,} \\
        Qionghai~Dai,~\IEEEmembership{Senior Member,~IEEE,}
        and~Tianyou~Chai,~\IEEEmembership{Fellow,~IEEE}
\IEEEcompsocitemizethanks{\IEEEcompsocthanksitem Gaochang Wu and Tianyou Chai are with the State Key Laboratory of Synthetical Automation for Process Industries, Northeastern University, Shenyang 110819, P. R. China.
\IEEEcompsocthanksitem Yebin Liu and Qionghai Dai are with the Broadband Network \& Digital Media Lab, Department of Automation, Tsinghua University, Beijing 100084, P. R. China.
\IEEEcompsocthanksitem Fang Lu is with Tsinghua-Berkeley Shenzhen Institute, Shenzhen 518055, P. R. China.}

\thanks{Corresponding Author: Yebin Liu, liuyebin@mail.tsinghua.edu.cn}
}

%
%

\markboth{Journal of \LaTeX\ Class Files,~Vol.~XX, No.~X, September~2017}%
{Shell \MakeLowercase{\textit{et al.}}: Bare Demo of IEEEtran.cls for Computer Society Journals}
%



\IEEEtitleabstractindextext{%
\begin{abstract}
In this paper, a novel convolutional neural network (CNN)-based framework is developed for light field reconstruction from a sparse set of views. We indicate that the reconstruction can be efficiently modeled as angular restoration on an epipolar plane image (EPI). The main problem in direct reconstruction on the EPI involves an information asymmetry between the spatial and angular dimensions, where the detailed portion in the angular dimensions is damaged by undersampling. Directly upsampling or super-resolving the light field in the angular dimensions causes ghosting effects. To suppress these ghosting effects, we contribute a novel ``blur-restoration-deblur'' framework. First, the ``blur'' step is applied to extract the low-frequency components of the light field in the spatial dimensions by convolving each EPI slice with a selected blur kernel. Then, the ``restoration'' step is implemented by a CNN, which is trained to restore the angular details of the EPI. Finally, we use a non-blind ``deblur'' operation to recover the spatial high frequencies suppressed by the EPI blur. We evaluate our approach on several datasets, including synthetic scenes, real-world scenes and challenging microscope light field data. We demonstrate the high performance and robustness of the proposed framework compared with state-of-the-art algorithms. We further show extended applications, including depth enhancement and interpolation for unstructured input. More importantly, a novel rendering approach is presented by combining the proposed framework and depth information to handle large disparities.
\end{abstract}

}

\maketitle

\IEEEdisplaynontitleabstractindextext

%
\IEEEpeerreviewmaketitle

\IEEEraisesectionheading{\section{Introduction}\label{sec:introduction}}
\IEEEPARstart{L}{ight} field imaging \cite{LFrendering,IhrkeRM16} is one of the most extensively used methods for capturing the 3D appearance of a scene. Rather than a limited collection of 2D images, a light field camera is able to collect not only the accumulated intensity at each pixel but also light rays from different directions. Early light field cameras, such as multi-camera arrays and light field gantries~\cite{CameraArray}, required expensive custom-made hardware or time-consuming capturing process.

In recent years, the introduction of commercial and industrial light field cameras, such as Lytro \cite{Lytro} and RayTrix \cite{RayTrix} has taken light field imaging into a new era. These plenoptic (light field) cameras are composed of microlens array and have the capacity of simultaneous capture. Unfortunately, due to the restricted sensor resolution, they must make a trade-off between spatial and angular resolution, i.e., one can obtain dense sampling images in the spatial dimensions but only sparse sampling in the angular (viewing angle) dimensions or vice versa.

\begin{figure}
\begin{center}
\includegraphics[width=1\linewidth]{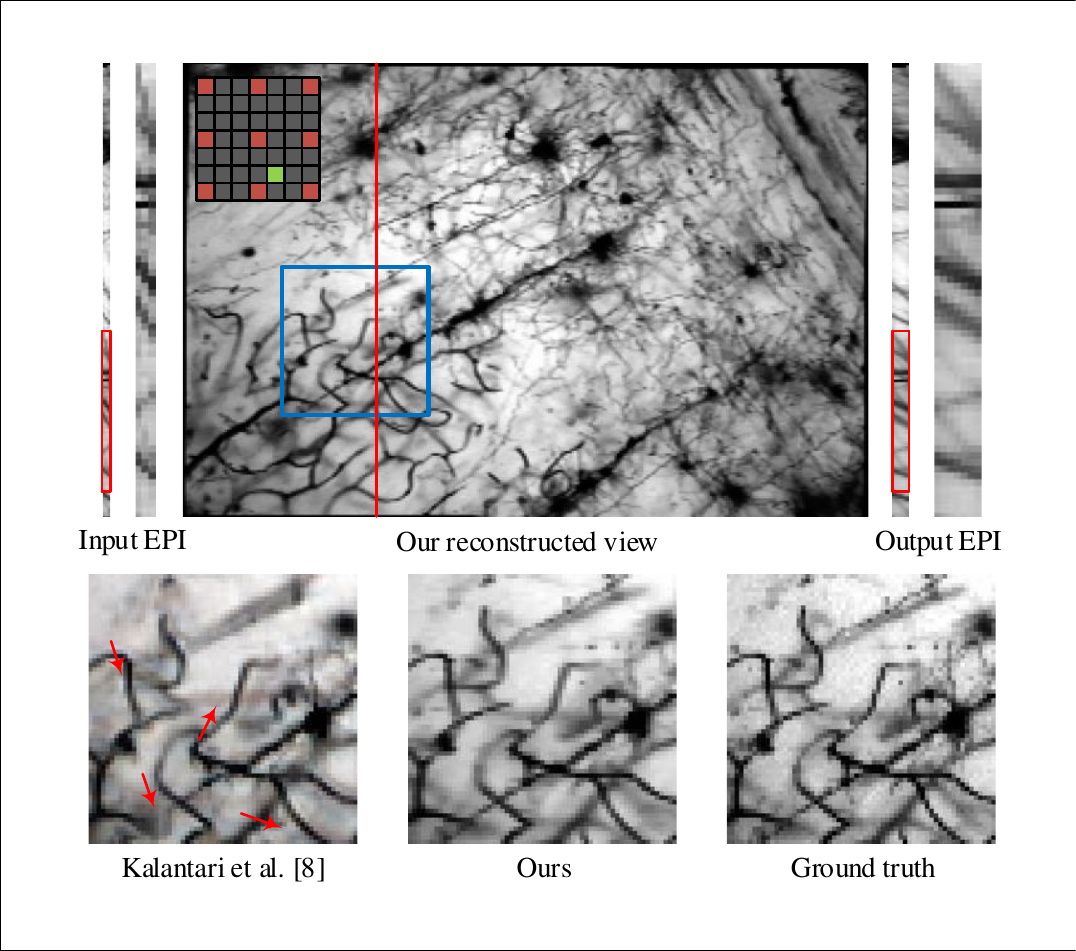}
\end{center}
\vspace{-4mm}
   \caption{Comparison of light field reconstruction results on Stanford microscope light field data \textit{Neurons} $20\times$ \cite{microLF} using $3\times3$ input views. The proposed learning-based EPI reconstruction produces better results in this challenging case.}
\label{fig:Teaser}
\vspace{-2mm}
\end{figure}

To solve this problem, various learning-based methods \cite{LFCNN,DeepStereo,DoubleCNN} have been proposed to super-resolve the light field in angular dimensions using a small set of views with high spatial resolution. In contrast to certain conventional studies \cite{Bayesian,LFfourier,binolf} that focus on novel view synthesis or reconstruction of the plenoptic function, learning-based methods train the network by directly minimizing the error between the synthesized view and the ground truth image. However, the network training is data dependent and cannot be easily transferred to data with different appearance properties, which limits the universal usage of the network. For example, microscopy scenes exhibit challenging structures and have very different appearances from the scenes in daily life, leading to undesirable results when training the network with macroscopic light fields (see Figure \ref{fig:Teaser}).


In this paper, we propose a novel learning-based framework to reconstruct high-angular-resolution light fields on an epipolar plane image (EPI). We indicate that by taking advantage of the special structure of the EPI, the light field reconstruction can be effectively modeled as learning-based angular detail restoration on this 2D structure. Compared with the sub-aperture images, the light field data share similar properties in the EPI domain, e.g., the EPI of microscope light field data (shown in Figure \ref{fig:Teaser}) has a similar structure as EPIs of macroscopic scenes (shown in Figure \ref{fig:Result2}).

We further indicate (see Sec. \ref{sec:formulation}) that the main problem in direct reconstruction on the EPI involves the information asymmetry between the spatial and angular dimension, where the high-frequency portion in the angular dimensions is damaged by undersampling. This information asymmetry will cause ghosting effects when the light field is directly super-resolved in the angular dimensions on the EPI \cite{LFGeometry,Chai2000}. To suppress the ghosting effect caused by this information asymmetry and simultaneously utilizing the spatial and angular information, we instead propose a ``blur-restoration-deblur'' framework on the EPI. First, in the ``blur'' step, we balance the information by extracting the spatial low-frequency components of the EPI. We implement this step by convolving each EPI slice with a selected blur kernel. Due to the coupling relationship between the spatial and angular dimensions in the EPI \cite{LFGeometry,Chai2000}, the ``blur'' step is equal to an anti-aliasing processing in the angular dimension. Then, in the ``restoration'' step, we apply a CNN to restore the angular detail of the EPI damaged by the undersampling. \wgcr{In this step, at least three views are used in each angular dimension to provide sufficient information for the restoration.} Finally, the ``deblur'' step is performed to recover the spatial detail suppressed by the EPI blur using a non-blind deblur operation.

Compared with state-of-the-art approaches that directly use sub-aperture images to generate novel views, our framework demonstrates better performance, especially on light fields containing complex occlusion regions, non-Lambertian surfaces and even challenging microscope scenes. Figure \ref{fig:Teaser} shows a comparison against a current state-of-the-art approach by Kalantari \textit{et al.} \cite{DoubleCNN} on the \textit{Neurons} $20\times$ case from the Stanford microscope light field data \cite{microLF}. The method by Kalantari \textit{et al.} \cite{DoubleCNN} results in blur in the occluded regions, whereas the proposed approach produces reasonable results even in this challenging case.

In summary, the contributions of this paper are as follows:

\begin{itemize}
\item{We take advantage of the clear texture structure of the EPI in the light field data and combine it with deep learning technique to super-resolve light field in the angular dimensions;}
\item{We reveal that the main problem in light field reconstruction using angularly sparsely input is the information asymmetry in the EPI. We therefore contribute a novel ``blur-restoration-deblur'' framework on EPI to address this problem. Extensive experiments on various light fields containing complex occlusion regions, non-Lambertian surfaces and even challenging microscope scenes have validated the efficiency of the proposed framework;}
\item{We show further applications including depth enhancement using reconstructed high angular resolution light field, unstructured light field super-resolution and depth assisted novel view rendering.}
\end{itemize}

A preliminary version of this paper appeared in \cite{EPICNN17}, which mainly introduced the ``blur-restoration-deblur'' framework for light field reconstruction. The present work mainly makes the following additional contributions compared with the preliminary version. First, we analyze the information asymmetry of the EPI and demonstrate the resulting ghosting effects in the Fourier domain. In addition, the efficacy of the proposed ``blur-restoration-deblur'' framework is also validated in the Fourier domain. Second, in addition to the application for depth enhancement shown in the preliminary version, we extend the proposed framework to additional applications including interpolation for unstructured input such as unstructured light fields. Third, we present a novel rendering scheme that seamlessly combines the proposed ``blur-restoration-deblur'' framework and depth information to address the interpolation with large disparity. This application inherits the rendering capability of handling large disparity from depth-image-based rendering techniques as well as the robustness to depth uncertainties, occlusion regions and non-Lambertian surfaces from the proposed framework. Fourth, an in-depth analysis on the merits of the learning-based reconstruction using EPIs is presented. The source code of this paper has been made public.

\section{Related Work}

The main obstacle in light field imaging is the trade-off between spatial and angular resolution due to limited sensor resolution. Super-resolution techniques to improve spatial and angular resolution have been studied by many researchers \cite{DBLP,PatchSR,Wanner,LFCNN,GuoYKLY16}. In this paper, we mainly focus on approaches for improving the angular resolution of the light field.


\subsection{Light field view synthesis}

Zhang \textit{et al.} \cite{binolf} proposed a phase-based approach using a micro-baseline stereo pair. They applied a disparity (depth)-assisted phase-based synthesis strategy to integrate the disparity information into the phase term when warping the input image to the novel view. However, their method was specifically designed for a micro-baseline stereo pair and causes artifacts in the occluded regions when extrapolating novel views. Zhang \textit{et al.} \cite{Plenopatch} described a patch-based approach for various light field editing tasks. In their work, the input depth map is decomposed into different depth layers and presented to the user to achieve the editing goals. However, these approaches rely heavily on the quality of the depth maps, which tends to fail in occluded, as well as glossy and specular, regions; thus, such approaches often fail to produce promising results.



Alternatively, some studies are based on sampling and consecutive reconstruction of the plenoptic function. For densely sampled light fields in which the disparity between the neighboring views does not exceed 1 pixel, novel views can be directly rendered by ray interpolation \cite{LFrendering}. For sparsely sampled light fields, a reconstruction in Fourier domain has been investigated in some studies. Levin and Durand \cite{Levin10} proposed a linear view synthesis approach using a dimensionality gap light field prior to synthesize novel views from a set of images sampled with a circular pattern. Shi \textit{et al.} \cite{LFfourier} considered light field reconstruction as an optimization for sparsity in the continuous Fourier domain. Their work sampled a small number of 1D viewpoint trajectories formed by a box and 2 diagonals to recover the full light field. However, these methods require the light field to be captured in a specific pattern, which limits its practical uses. Didyk \textit{et al.} \cite{Didyk15} used the phase information from a complex steerable pyramid decomposition to synthesize novel views with a small displacement; for large displacements, only low-frequency components can be reconstructed.

\subsection{Light field EPI structure}

By taking advantage of the EPI structure, Wanner and Goldluecke \cite{Wanner} employed the structure tensor of an EPI to perform fast and robust local disparity estimation; then, a TV-$L^1$ optimization scheme is applied to smooth the local result. Based on Wanner and Goldluecke's work, a certainty map was proposed to enforce visibility constraints on the initial estimated depth map in \cite{ContinuousDepth}. However, when implementing angular super-resolution, Wanner and Goldluecke \cite{Wanner} fell back into the sub-aperture image space and warped the input images to synthesize novel views based on the disparity information. In contrast, Vagharshakyan \textit{et al.} \cite{Shearlet} considered the angular super-resolution as an inpainting problem on the EPI, and the angular aliasing could be suppressed in the Fourier domain. They therefore utilized an adapted discrete shearlet transform to reconstruct the light field from a sparse sampled light field. However, the reconstruction exhibited poor quality in the border regions, resulting in a reduction of angular extent. Moreover, high-frequency components in the EPI are also lost when using a discrete shearlet to suppress high-frequency leakage caused by angular aliasing.

\subsection{Learning-based methods}

Recently, learning-based techniques have been explored for light field reconstruction. Cho \textit{et al.} \cite{Cho13} adopted a sparse-coding-based (SC) method to reconstruct light fields using raw data. They generate image pairs using Barycentric interpolation. Yoon \textit{et al.} \cite{LFCNN} trained a deep neural network for spatial and angular super-resolution. However, the network uses pairs of images to generate a novel view between them; thus, the network underused the potential of the full light field. Wang \textit{et al.} \cite{LFrecon} proposed several CNN architectures, one of which was developed for the EPI slices; however, the network is designed for material recognition, which is different from the EPI restoration task.

In addition, some studies on maximizing the quality of synthesized views that are based on CNNs have been presented. Flynn \textit{et al.} \cite{DeepStereo} proposed a deep learning method to synthesize novel views using a sequence of images with wide baselines. Kalantari \textit{et al.} \cite{DoubleCNN} used two sequential convolutional neural networks to model depth and color estimation simultaneously by minimizing the error between synthesized views and ground truth images. \wgcr{However, in that study, the network is trained using a fixed sampling pattern (four corner views), which greatly limits its generalizability. Although the proposed framework applies at least three views in each angular dimension (limited by the initial bicubic interpolation), our approach is suitable for inputs with different degrees of sparsity.} In addition, the approach results in ghosting artifacts in the occluded regions and fails to handle certain challenging cases.

\begin{figure}
\begin{center}
\includegraphics[width=1\linewidth]{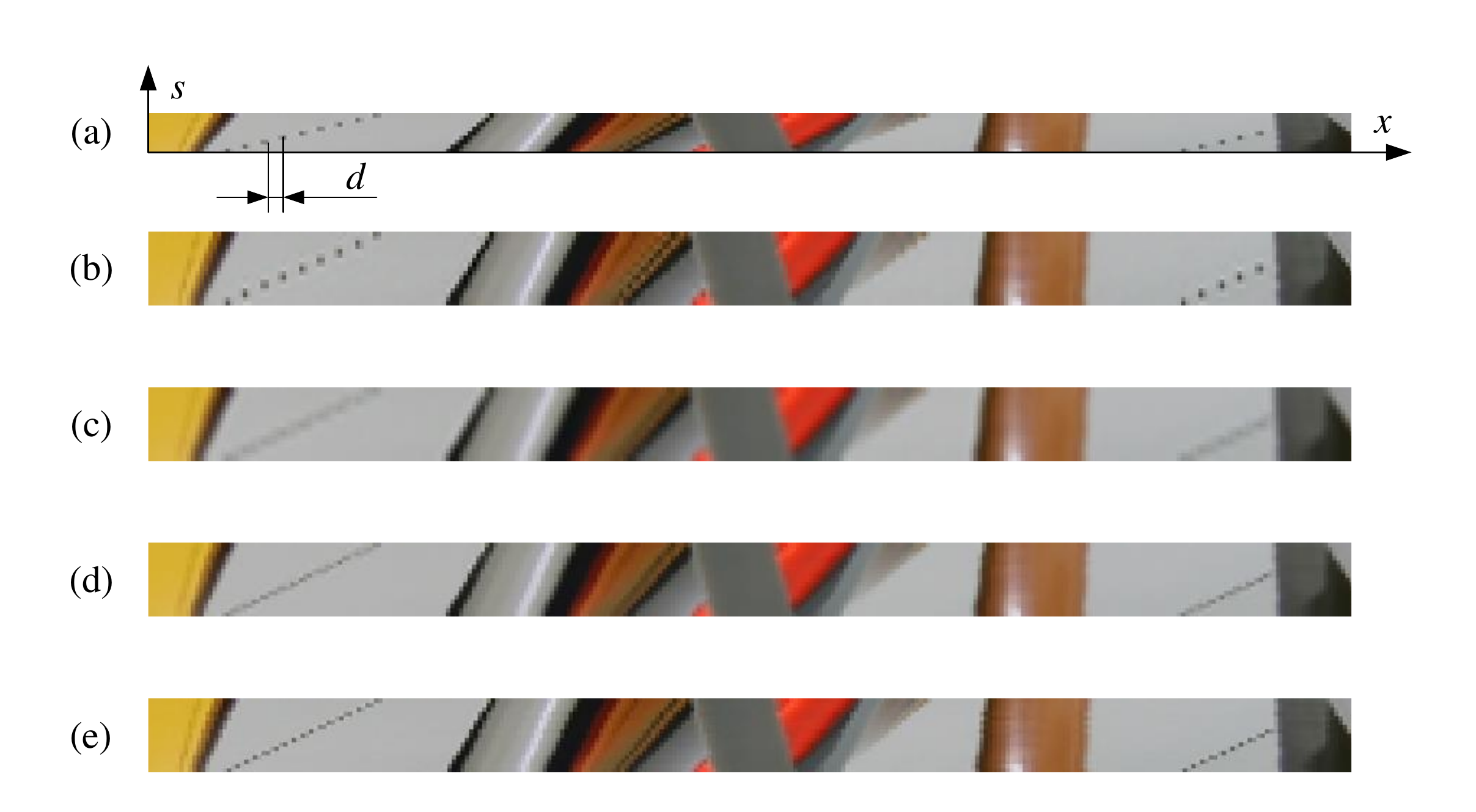}
\end{center}
\vspace{-4mm}
   \caption{An illustration of EPI upsampling results. (a) The input low-angular-resolution EPI, where $d=4$ pixels is the disparity between the neighboring views. See the angular aliasing effects in the low-angular-resolution EPI; (b) Straightforward CNN-based angular super-resolution will cause aliasing in the EPI, leading to ghosting effects in the reconstructed light field; (c) The result after using EPI blur (on the spatial dimension) and bicubic interpolation (on the angular dimension); (d) The final high-angular-resolution EPI produced by the proposed framework; (e) The ground truth high-angular-resolution EPI.}
\label{fig:LowF}
\vspace{-2mm}
\end{figure}

\section{Problem Analysis and Formulation}
\label{sec:formulation}

For a 4D light field $L(x,y,s,t)$, where $x$ and $y$ are the spatial dimensions and $s$ and $t$ are the angular dimensions, a 2D slice can be acquired by gathering horizontal lines with fixed $y^*$ along a constant camera coordinate $t^*$, denoted as $E_{y^*,t^*}(x,s)$. This 2D slice is called an epipolar plane image (EPI). The low-angular-resolution EPI $E_L$ is a down-sampled version of the high-angular-resolution EPI $E_H$:
\vspace{-1mm}
\begin{equation}
\textbf{E}_L=\textbf{E}_H \downarrow,
\end{equation}
where $\downarrow$ denotes the down-sampling operation. Our task is to find an inverse operation $F$ that can minimize the error between the reconstructed EPI and the original high-angular-resolution EPI:
\vspace{-1mm}
\begin{equation}
\hat{F}=\mathop {\min }\limits_F||\textbf{E}_H-F(\textbf{E}_L)||.
\end{equation}

For a densely sampled light field, where the disparity between the neighboring views does not exceed 1 pixel, the angular sampling rate satisfies the Nyquist sampling criterion (the details of this deduction can be found in \cite{LFGeometry,Chai2000}). One can reconstruct such a light field based on the plenoptic function; however, for light fields sampled under the Nyquist sampling rate in the angular dimensions, the disparity is always larger than 1 pixel (see Figure \ref{fig:LowF}(a)). This undersampling of the light field destroys the high-frequency components in the angular dimension, whereas the spatial information is complete. This information asymmetry between the angular and spatial information causes aliasing effects in the EPI, which will be further aggravated if the angular resolution is directly super-resolved (see Figure \ref{fig:LowF}(b)). For example, in Figure \ref{fig:LowF}(e), the black line in the ground truth EPI is continuous, whereas the directly super-resolved EPI in Figure \ref{fig:LowF}(b) cannot reconstruct the line in this case. The aliasing in the EPI will lead to ghosting effects in the reconstructed light field. Note that this information asymmetry will always occur when the disparity between the neighboring views is larger than 1 pixel. A more detailed analysis of angular aliasing effects in spatial domain can be found in \cite{Xiao2014Aliasing}.

\begin{figure}
\begin{center}
\includegraphics[width=1\linewidth]{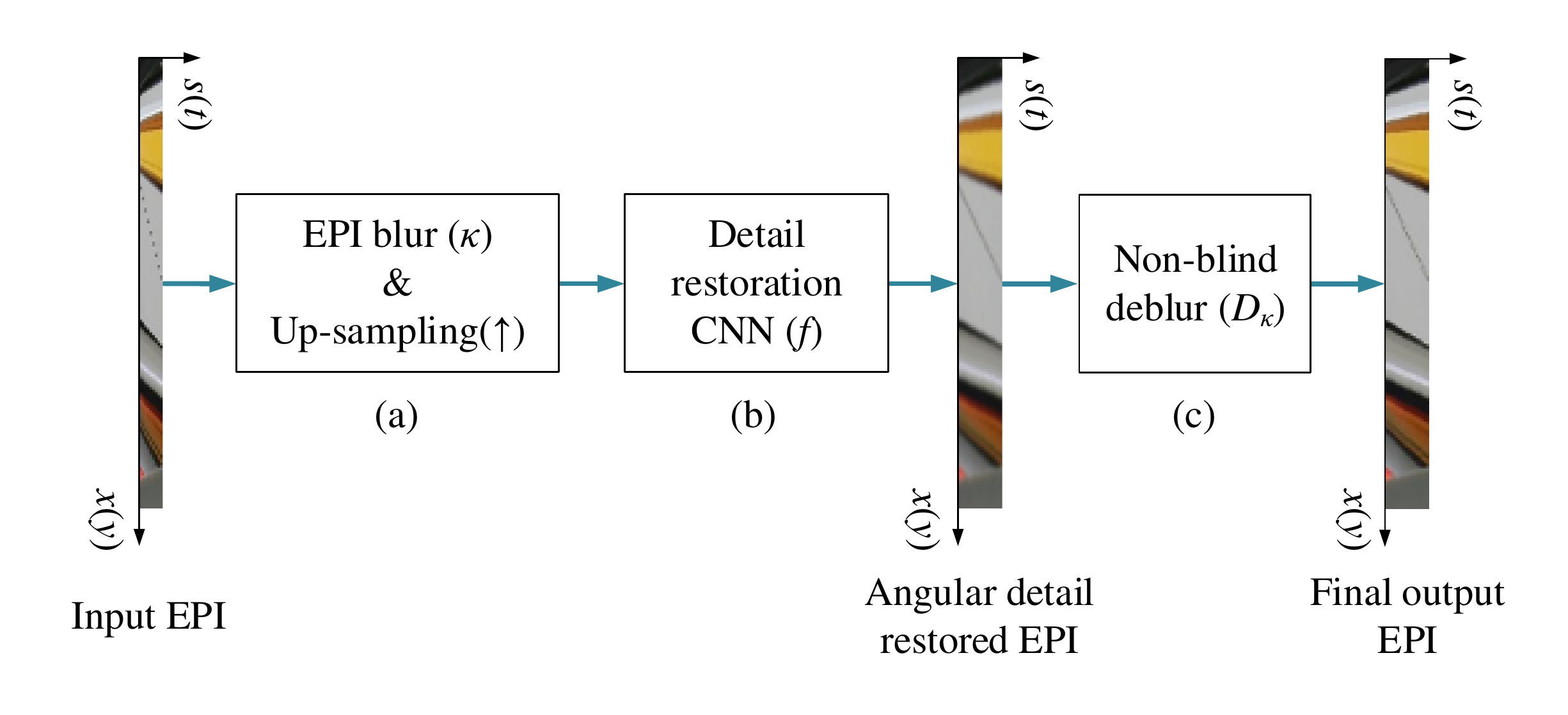}
\end{center}
\vspace{-4mm}
   \caption{The proposed ``blur-restoration-deblur'' framework for light field reconstruction on an EPI.}
\label{fig:Framework}
\end{figure}

To ensure information symmetry between the spatial and angular dimensions of the EPI, one can decrease the spatial resolution of the light field to an appropriate level. However, it is then difficult to recover the novel views with the original spatial quality, especially when a large downsampling rate has to be used, as shown in the case in Figure \ref{fig:LowF} (a). Rather than decreasing the spatial resolution of the light field, we extract the low-frequency information by convolving the EPI with a 1D blur kernel in the spatial dimension. Due to the coupling relationship between the spatial and angular dimension \cite{LFGeometry,Chai2000}, the ``blur'' step equals an anti-aliasing processing in the angular dimension. In addition, because the kernel is predesigned, the spatial details can be easily recovered using a non-blind deblur operation after the angular restoration processing. Figure \ref{fig:LowF}(c) shows the blurred and upsampled result of the sparse sampled EPI in Figure \ref{fig:LowF}(a). We now reformulate the reconstruction of the EPI $\textbf{E}_L$ as follows:
\begin{equation}
f=\mathop {\arg\min}\limits_f||\textbf{E}_H-D_{\kappa}f((\textbf{E}_L* \kappa)\uparrow)||,
\end{equation}
where $*$ is the convolution operator, $\kappa$ is the blur kernel, $\uparrow$ is a bicubic interpolation operation that upsamples the EPI to the desired angular resolution, $f$ represents an operation that recovers the high-frequency detail in the angular dimension, and $D_\kappa$ is a non-blind deblur operator that uses the kernel $\kappa$ to recover the spatial detail of the EPI suppressed by the EPI blur. In our paper, we model the restoration process $f$ with a CNN to learn a mapping between the blurred low-angular-resolution EPI and the blurred high-angular-resolution EPI. The final reconstructed high angular resolution EPI $\hat{\textbf{E}}_H=D_{\kappa}f((\textbf{E}_L* \kappa)\uparrow$).

\section{Proposed Framework}

\subsection{Overview}

The EPI is the building block of a light field and contains both the angular and spatial information. We take advantage of this characteristic to model the reconstruction of the sparsely sampled light field as learning-based angular information restoration on the EPI (Equation 3). To avoid information asymmetry, we propose a ``blur-restoration-deblur'' framework, which is shown in Figure \ref{fig:Framework}.

In the first ``blur'' step, we extract the spatial low-frequency information of the EPI using EPI blur (see Figure \ref{fig:Framework}(a)). We then upsample the EPI to the desired angular resolution using bicubic interpolation in the angular dimension. Then, in the ``restoration'' step, we apply a CNN to restore the details of the EPI in the angular dimension (see Figure \ref{fig:Framework}(b)). The network architecture is similar to that in \cite{SRCNN}. The main difference is that we apply a residual-learning method to predict only the angular details of the EPI. The network details are presented in Sec. \ref{sec:CNN}. In the final ``deblur'' step, the spatial details of the EPI are recovered through a non-blind deblur operation \cite{deblur} (see Figure \ref{fig:Framework}(c)), and the output EPIs are applied to reconstruct the final high-angular-resolution light field. It should be noted that the CNN is trained to restore the angular details that are damaged by the undersampling of the light field rather than the spatial details suppressed by the EPI blur. An alternative approach is to model the deblur operation into the CNN; however, using that approach, the network will inevitably be deeper, slower to converge and harder to tune, making it more difficult to produce good results. Comparatively, the non-blind deblur is substantially more suitable to the task because the kernel is known.

\begin{figure}
\begin{center}
\includegraphics[width=0.95\linewidth]{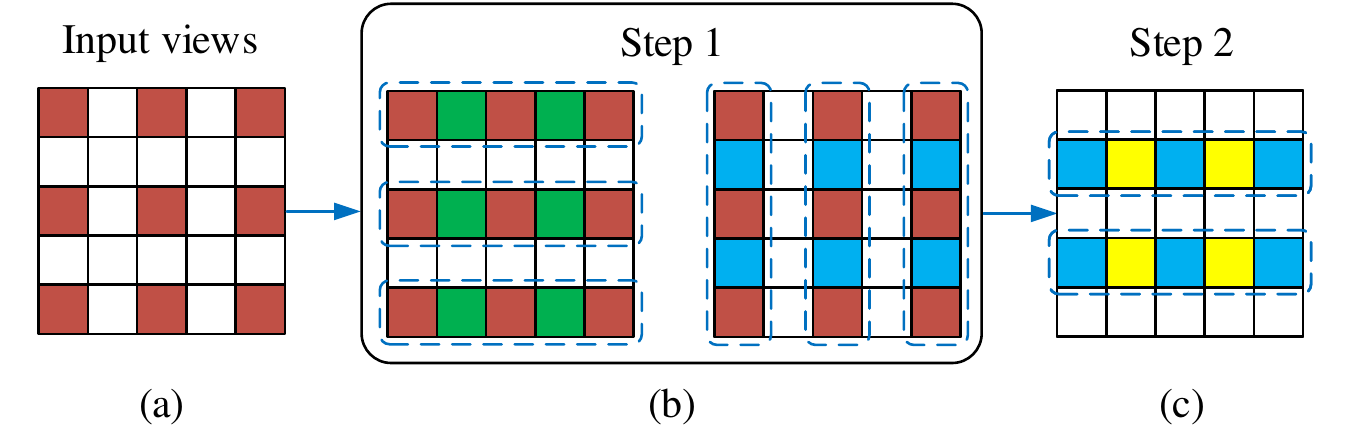}
\end{center}
\vspace{-4mm}
   \caption{\wgcr{Hierarchical reconstruction of the full light field. (a) The input light field is composed of the images marked in red; (b) In Step 1, the EPIs from the horizontal views (in the left dashed boxes) are used to generate the novel views marked in green, and the EPIs from the vertical views (in the right dashed boxes) are used to generate the novel views marked in blue. (c) In Step 2, the views generated from the Step 1 (in the dashed boxes) are used to produce the rest of views (marked in yellow).}}
\label{fig:LFs2d}
\end{figure}

\wgcr{To obtain the full light field using the sparsely sampled light field, we adopt a hierarchical reconstruction strategy (as illustrated in Figure \ref{fig:LFs2d}). In Step 1 (Figure \ref{fig:LFs2d}(b)), the input light field (views marked in red in Figure \ref{fig:LFs2d}(a)) is used to produce a part of novel views. In this step, the EPIs from the horizontal views are used to generate the novel views marked in green, and the EPIs from the vertical views are used to generate the novel views marked in blue. In Step 2 (Figure \ref{fig:LFs2d}(c)), the images generated from the Step 1 (marked in blue) are applied to produce the rest of views (marked in yellow) in the final high angular resolution light field. Alternatively, one can use the images marked in green (in Figure \ref{fig:LFs2d}(b)) to produce the the rest of the views; however, we empirically found that the selection of the input images in the Step 2 has a negligible influence on the final results. We therefore apply the former method in our experiments. The EPI reconstruction in each step is performed by using the ``blur-restoration-deblur'' framework.}

\subsection{Low-frequency extraction based on EPI blur}

To extract the low frequencies of the EPI from only the spatial dimension, we define the blur kernel in 1D space rather than defining a 2D image blur kernel. The following candidates are considered when extracting the low-frequency part of the EPIs: the sinc function, the spatial representation of a Butterworth low-pass filter of order 2 and the Gaussian function. The spatial representations of the filters are as follows:
\begin{equation}
\begin{split}
\kappa_s(x)&=c_1\rm{sinc}\it(x/(\rm2\it|\sigma|)),\\
\kappa_b(x)&=c_2e^{-|x/\sigma|}(\cos(|x/\sigma|)+\sin(|x/\sigma|)),\\
\kappa_g(x)&=c_3e^{-x^2/(2\sigma^2)},
\end{split}
\end{equation}
where $c_1$, $c_2$ and $c_3$ are scale parameters, and $\sigma$ is a shape parameter. In our paper, the kernels are discretized at the integer coordinate and limited to a finite window, i.e., $x\in [-4\sigma,4\sigma]$. The kernel size is determined by the largest disparity (e.g., for the light field with the largest disparity of 4 pixels, the shape parameter $\sigma=1.5$, and the kernel size is 13). The scale parameters are used to normalize the kernels.

We evaluate these three kernels based on the following two principles: the final deblurred result must show visual coherency with the ground truth EPI, and the mean squared error (MSE) between the blurred low-angular-resolution EPI and the blurred ground truth EPI is as minimal as possible:
\begin{equation}
\hat{\kappa}=\mathop {\min }\limits_{\kappa}\frac{1}{n}{\sum_{i=1}^n{||(\textbf{E}_L^{(i)}* \kappa)\uparrow-\textbf{E}^{(i)}* \kappa||^2}},
\end{equation}
where $i$ is the index of the EPIs, $n$ is the number of EPIs, $\textbf{E}_L$ represents the low-angular-resolution EPIs, and $\textbf{E}$ represents the ground truth high-angular-resolution EPIs. We evaluate the kernels on the Stanford Light Field Archive \cite{StanfordLFdatasets}, and the errors between the processed (blurred and upsampled) EPIs and the blurred ground truth EPIs are $0.153$, $0.089$ and $0.061$ for the sinc, Butterworth and Gaussian kernels, respectively. The sinc function represents an ideal low-pass filter in the spatial dimension, and the low frequencies can pass through the filter without distortion. However, this ideal low-pass filter causes ringing artifacts in the EPIs. The Butterworth kernel generates imperceptible ringing artifacts, whereas the Gaussian ensures that no ringing artifacts exist. Based on this observation and the numerical evaluation, the Gaussian function is selected to be the kernel for the EPI blur.

\subsection{Detail restoration based on CNN}
\label{sec:CNN}

For CNN-based image restoration, Dong \textit{et al.} \cite{SRCNN} proposed a network for single image super-resolution named SRCNN, in which a high-resolution image is predicted from a given low-resolution image. Kim \textit{et al.} \cite{deepSRCNN} improved on that work using a residual network with a deeper structure. Inspired by those pioneers, we design a residual network with three convolution layers to restore the angular detail of the EPIs.

\subsubsection{CNN architecture}

\begin{figure}
\begin{center}
\includegraphics[width=1\linewidth]{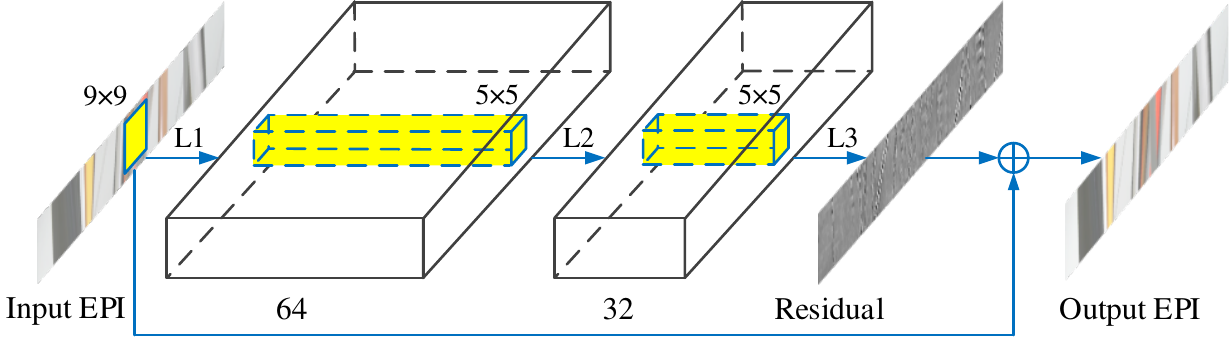}
\end{center}
\vspace{-4mm}
   \caption{The proposed detail restoration network is composed of three layers. The first and second layers are followed by a rectified linear unit (ReLU). The final output of the network is the sum of the predicted residual (detail) and the input.}
\label{fig:CNN}
\end{figure}

The architecture of the detail restoration network is outlined in Figure \ref{fig:CNN}. Consider an EPI that is convolved with the blur kernel and up-sampled to the desired angular resolution, denoted as $\textbf{E}'_L$ for short. The desired output EPI $f(\textbf{E}'_L)$ is then the sum of the input $\textbf{E}'_L$ and the predicted residual $\mathcal{R}(\textbf{E}'_L)$:
\begin{equation}
f(\textbf{E}'_L)=\textbf{E}'_L+\mathcal{R}(\textbf{E}'_L).
\end{equation}
The network for the residual prediction consists of three convolution layers. The first layer contains 64 filters of size $1\times9\times9$, where each filter operates on a $9\times9$ spatial region across 64 channels (feature maps) and is used for feature extraction. The second layer contains 32 filters of size $64\times5\times5$ and is used for non-linear mapping. The last layer contains 1 filter of size $32\times5\times5$ and is used for detail reconstruction. Both the first and second layers are followed by a rectified linear unit (ReLU). Due to the limited angular information of the light field used as the training dataset, we pad the data with zeros before every convolution operation to maintain the input and output as the same size.

We apply this residual learning method for the following reasons. First, the undersampling in the angular dimension damages the high-frequency portion (detail) of the EPIs; thus, only those details need to be restored. Second, extracting these details prevents the network from having to consider the low-frequency part, which would be a waste of time and result in reduced accuracy.

\subsubsection{Training details}

The desired residuals are $\textbf{R}=\textbf{E}'-\textbf{E}'_L$, where $\textbf{E}'$ are the blurred ground truth EPIs and $\textbf{E}'_L$ are the blurred and interpolated low-angular-resolution EPIs. Our goal is to minimize the mean squared error $\frac{1}{2}||\textbf{E}'-f(\textbf{E}'_L)||^2$. However, due to the residual network that we use, the loss function is now formulated as follows:
\begin{equation}
L=\frac{1}{n}\sum_{i=1}^{n}||\textbf{R}^{(i)}-\mathcal{R}(\textbf{E}_L'^{(i)})||^2,
\end{equation}
where $n$ is the number of training EPIs. The output of the network $\mathcal{R}(\textbf{E}'_L)$ represents the restored details, which must be added back to the input EPI $\textbf{E}'_L$ to obtain the final high-angular-resolution EPI $f(\textbf{E}'_L)$.

We use the Stanford Light Field Archive \cite{StanfordLFdatasets} (captured using a gantry system) as the training data. The blurred ground truth EPIs are decomposed into sub-EPIs of size $17\times17$ with stride of 14. To avoid overfitting, we adopted data augmentation techniques \cite{dataAugment1,dataAugment2} that include flipping, downsampling the spatial resolution of the light field as well as adding Gaussian noise. To avoid the limitations of a fixed angular up-sampling factor, we use a scale augmentation technique. Specifically, we downsample some EPIs with a small angular extent by a factor of 4 and the desired output EPIs by a factor of 2; then, we upsample them to the original resolution. The network is trained by using the datasets downsampled by both a factor of 2 and a factor of 4. We use the cascade of the network for the EPIs that are required to be up-sampled by a factor of 4. In practice, we extract more than $8e6$ examples, which is sufficient for the training. We select the mini-batches of size 64 as a trade-off between speed and convergence.

In the paper, we followed the conventional methods of image super-resolution to transform the EPIs into YCbCr space: only the Y channel (i.e., the luminance channel) is applied to the network. This is because the other two channels are blurrier than the Y channel and, thus are less useful in the restoration \cite{SRCNN}.

\begin{figure}
\begin{center}
\includegraphics[width=0.95\linewidth]{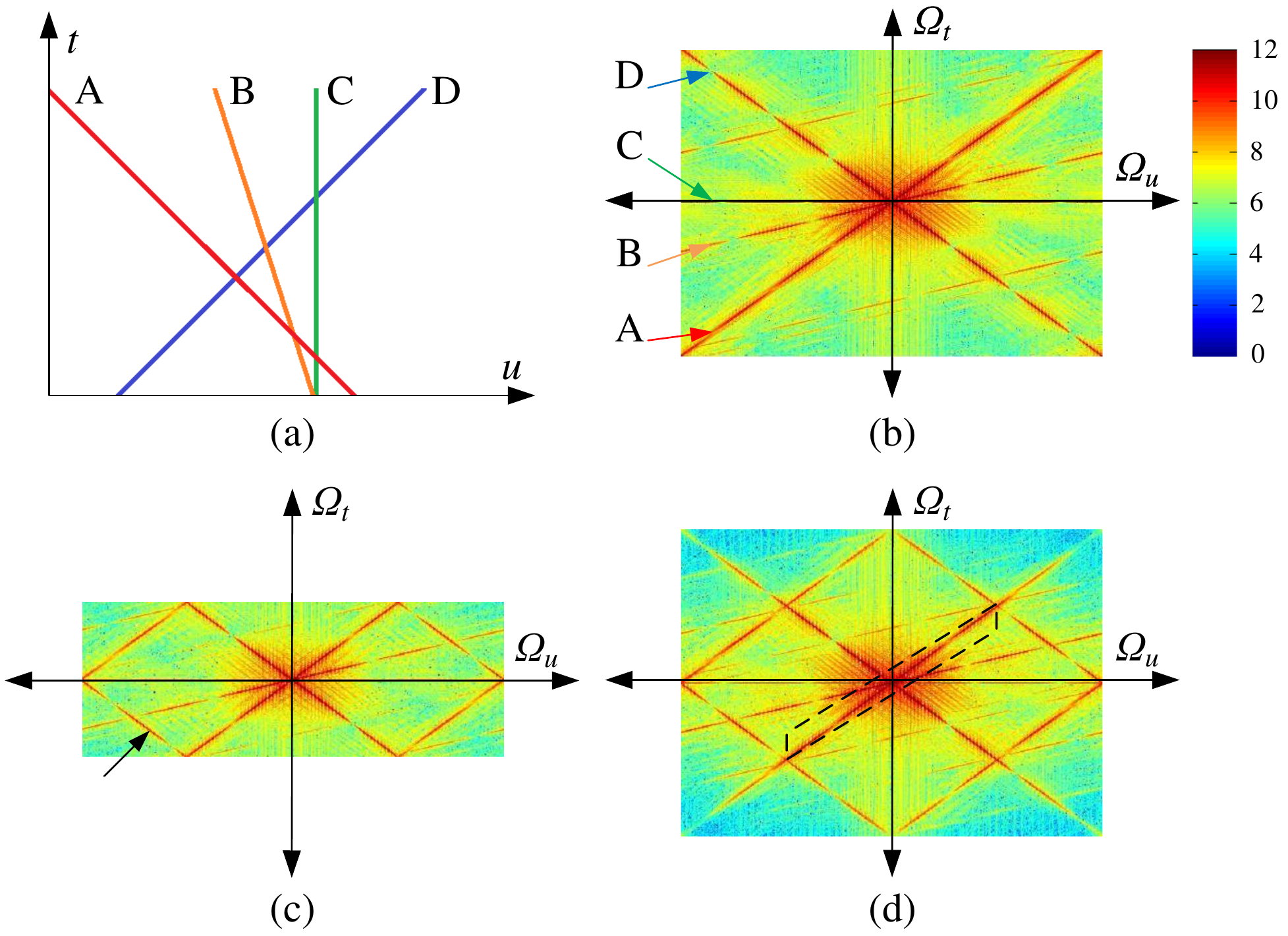}
\end{center}
\vspace{-4mm}
\caption{\wgcb{(a) A densely sampled EPI that contains four lines of different slopes. The disparity is no greater than 1 pixel. (b) The Fourier spectrum of the EPI in (a), where the lines in (a) are marked with arrows in their corresponding colors. Note that the Fourier spectrum of the green line is occluded by the $\Omega_u$-axis. (c) The light field is downsampled in the angular dimensions, producing an angularly undersampled EPI. The undersampling generates copies of the Fourier spectrum, where one copy is indicated by a black arrow. (d) Direct CNN-based super-resolution causes high-frequency leakage from the copies. The band-limited filter shown in the black dashed box can only reconstruct light field to a certain depth. The color bar on the right side of (b) shows the power range of the Fourier spectrum after taking the logarithm.}}
\label{fig:FA1}
\end{figure}

To improve the convergence speed, we adjust the learning rate consistently with the increasing of the training iteration. The number of training iterations is $8 \times10^5$. The learning rate is set to 0.01 initially and decreased by a factor of 10 every $0.25\times10^5$ iterations. When the number of training iterations is $5.0\times10^5$ , the learning rate is decreased to 0.0001 in two reduction steps. We initialize the filter weight of each layer using a Gaussian distribution with zero mean and standard deviation $1e^{-3}$. The momentum parameter is set to 0.9. Training takes approximately 12 hours on a GTX 960 GPU (Intel CPU E3-1231 running at 3.40 GHz with 32 GB of memory). The training model is implemented using the \textit{Caffe} package \cite{Caffe}.

\section{\wgcb{Fourier Analysis}}\label{Sec:FA}

\wgcb{In this section, we further analyze the proposed ``blur-restoration-deblur'' framework in the Fourier domain. In the following, we will show how angular undersampling influences the Fourier spectrum of an EPI \cite{wu2017light,Stewart03}, explain why we apply the ``blur'' step to remove the spatial high-frequency components of an EPI, and demonstrate the performance of the proposed framework in the Fourier domain.}

\wgcb{Consider a simple scene composed of four points located at different depths. For an appropriately sampled light field, the resulting EPI contains four lines of different slopes, where each of the disparities does not exceed 1 pixel (as shown in Figure \ref{fig:FA1}(a)). Figure \ref{fig:FA1}(b) shows the Fourier spectrum of the EPI, where the angles of intersection with the $\Omega_u$-axis are determined by the depths of the objects in the scene. In Figure \ref{fig:FA1}(b), we mark the Fourier spectrum of each line in Figure \ref{fig:FA1}(a) with an arrow in its corresponding color.}

\begin{figure}
\begin{center}
\includegraphics[width=0.95\linewidth]{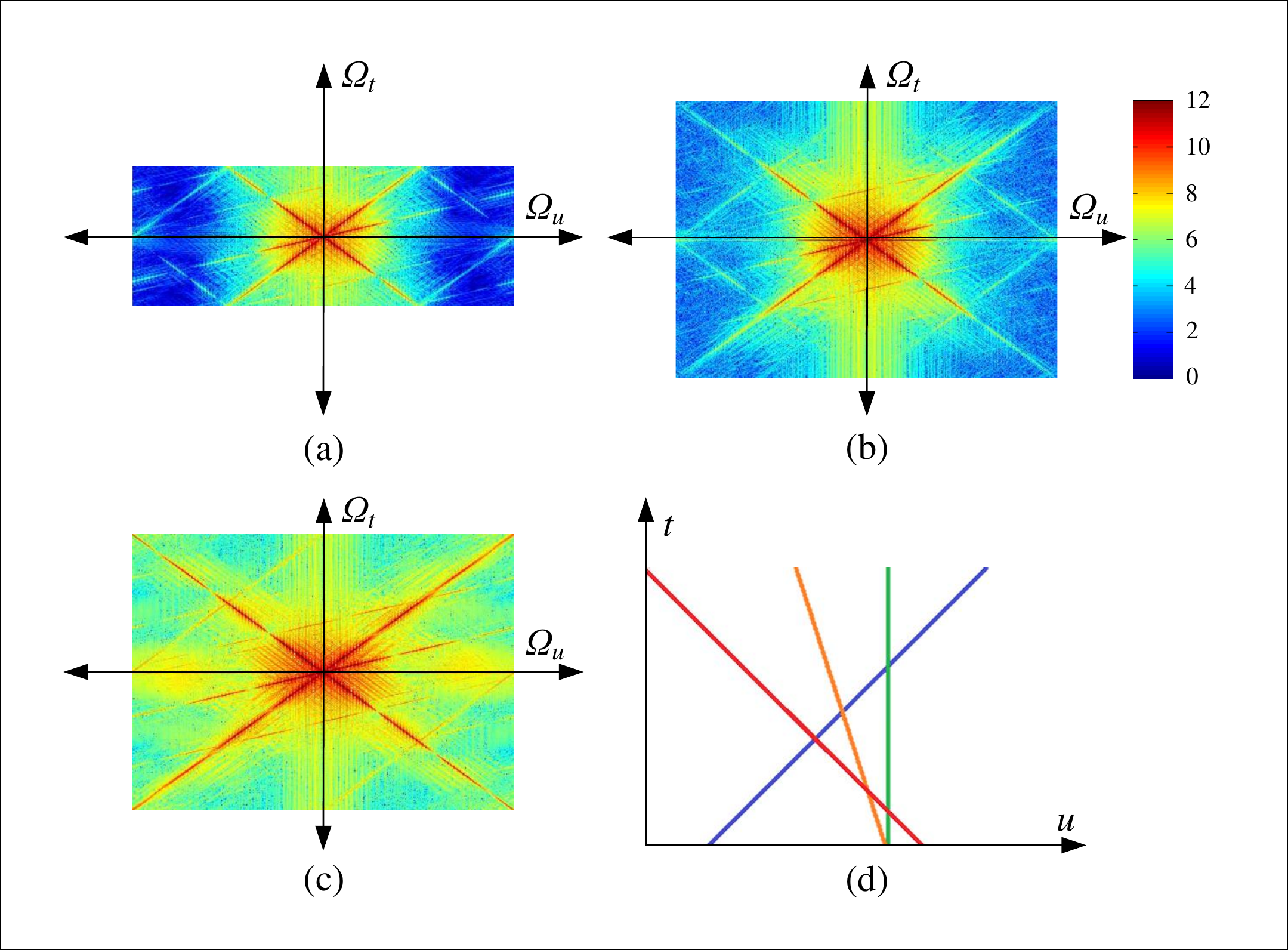}
\end{center}
\vspace{-4mm}
\caption{\wgcb{(a) The Fourier spectrum of the undersampled EPI after the ``blur'' step, where the high-frequency copies are efficiently suppressed. (b) The Fourier spectrum of the EPI after the ``restoration'' step. Compared with the Fourier spectrum in (a), the high-frequency components are restored, while the high copies that lead to aliasing are remained unchanged. (c) The Fourier spectrum of the EPI after being processed by the entire framework. (d) The super-resolved EPI produced by the proposed framework.}}
\label{fig:FA2}
\end{figure}

\wgcb{We simulate the sparsely sampled light field in the angular domain by downsampling the light field in the angular dimensions, generating an angularly undersampled EPI whose disparity falls outside the one-pixel range. The sampling weakly influences on the point with a small disparity. However, for points with a large disparity, the sampling destroys high-frequency details in the angular dimension, producing copies of the Fourier spectrum, as shown in Figure \ref{fig:FA1}(c). Straightforward upsampling or CNN-based super-resolution will cause high-frequency leakage from the other copies \cite{Isaksen2000} (shown in Figure \ref{fig:FA1}(d)), which further aggravates the angular aliasing in the EPI, as shown in Figure \ref{fig:LowF}(b).}

\begin{figure*}
\begin{center}
\includegraphics[width=1\linewidth]{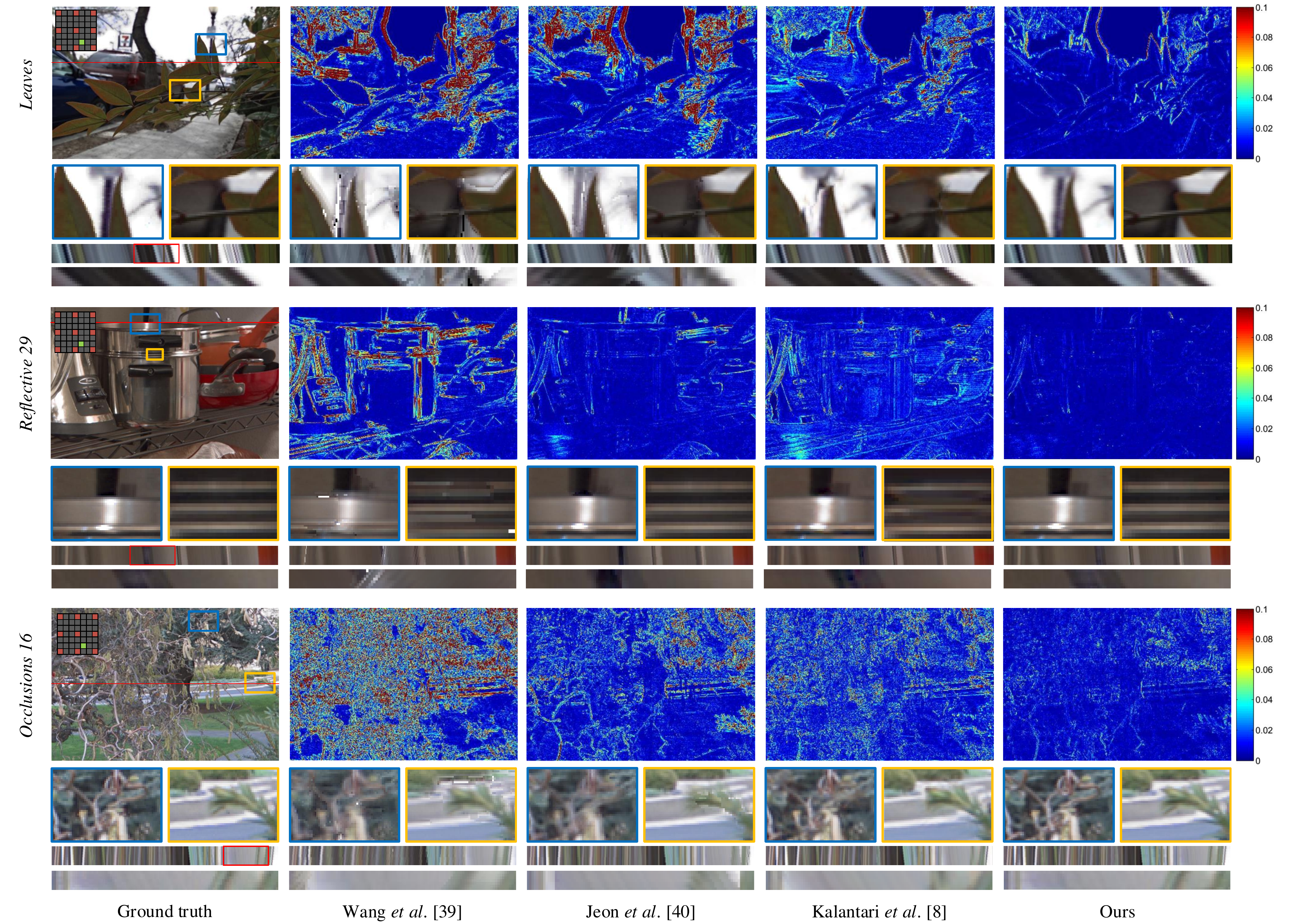}
\end{center}
\vspace{-4mm}
   \caption{Comparison of the proposed approach against other methods on the real-world scenes. The results show the ground truth images, error maps of the synthesized results in the Y channel, close-up versions of the image portions in the blue and yellow boxes, and the EPIs located at the red line shown in the ground truth view. The EPIs are upsampled to an appropriate scale in the angular dimension for better viewing. The lowest image in each block shows a close-up of the portion of the EPIs in the red box.}
\label{fig:Result2}
\end{figure*}

\wgcb{To overcome the aliasing in the EPI, Stewart \textit{et al.} \cite{Stewart03} applied a band-limited filter to reconstruct a light field in the Fourier domain. The filter preserves certain frequency components and simultaneously removes high-frequency leakage by changing the shape of the filter (shown in the black dashed box in Figure \ref{fig:FA1}(d)). However, Liang and Ramamoorthi \cite{Liang15} indicated that the filter is depth dependent, and simply reshaping the filter cannot be used to reconstruct light field in all depthes of field. Instead, in the proposed ``blur-restoration-deblur'' framework, we adopt a novel learning-based method which is able to reconstruct an EPI in a large disparity range without introducing aliasing effects.}

\wgcb{Specifically, we first balance the information between the spatial and angular information by applying a ``blur'' step. Unlike the band-limited filter described above, we use a simple 1D Gaussian kernel whose kernel size depends on the highest depth (disparity) of the light field. This ``blur'' step removes the high-frequency components in the spatial dimension. Figure \ref{fig:FA2}(a) shows the Fourier spectrum of the blurred undersampled EPI. Compared with the Fourier spectrum of the undersampled EPI shown in Figure \ref{fig:FA1}(c), the copies lying in the high-frequency regions are effectively suppressed. The blurred EPI is upsampled to the desired angular resolution using bicubic interpolation.}

\wgcb{Then, the CNN-based ``restoration'' step is performed to restore the angular details. Figure \ref{fig:FA2}(b) shows the Fourier spectrum of the EPI after the ``restoration'' step. From the perspective of the Fourier spectrum, the CNN is trained to restore the high-frequency components rather than the high-frequency copies that lead to aliasing.}

\wgcb{The ``deblur'' step using the selected Gaussian kernel is adopted to recover the high-frequency components in the spatial dimension, which is an inverse operation with respect to the ``blur'' step. Figure \ref{fig:FA2}(c) and (d) show the Fourier spectrum and the EPI, respectively, after being processed by the entire ``blur-restoration-deblur'' framework. Due to the removal of the high-frequency copies in the Fourier spectrum, the EPI is finally super-resolved without aliasing.}

\section{Evaluation}

\begin{figure*}
\begin{center}
\includegraphics[width=1\linewidth]{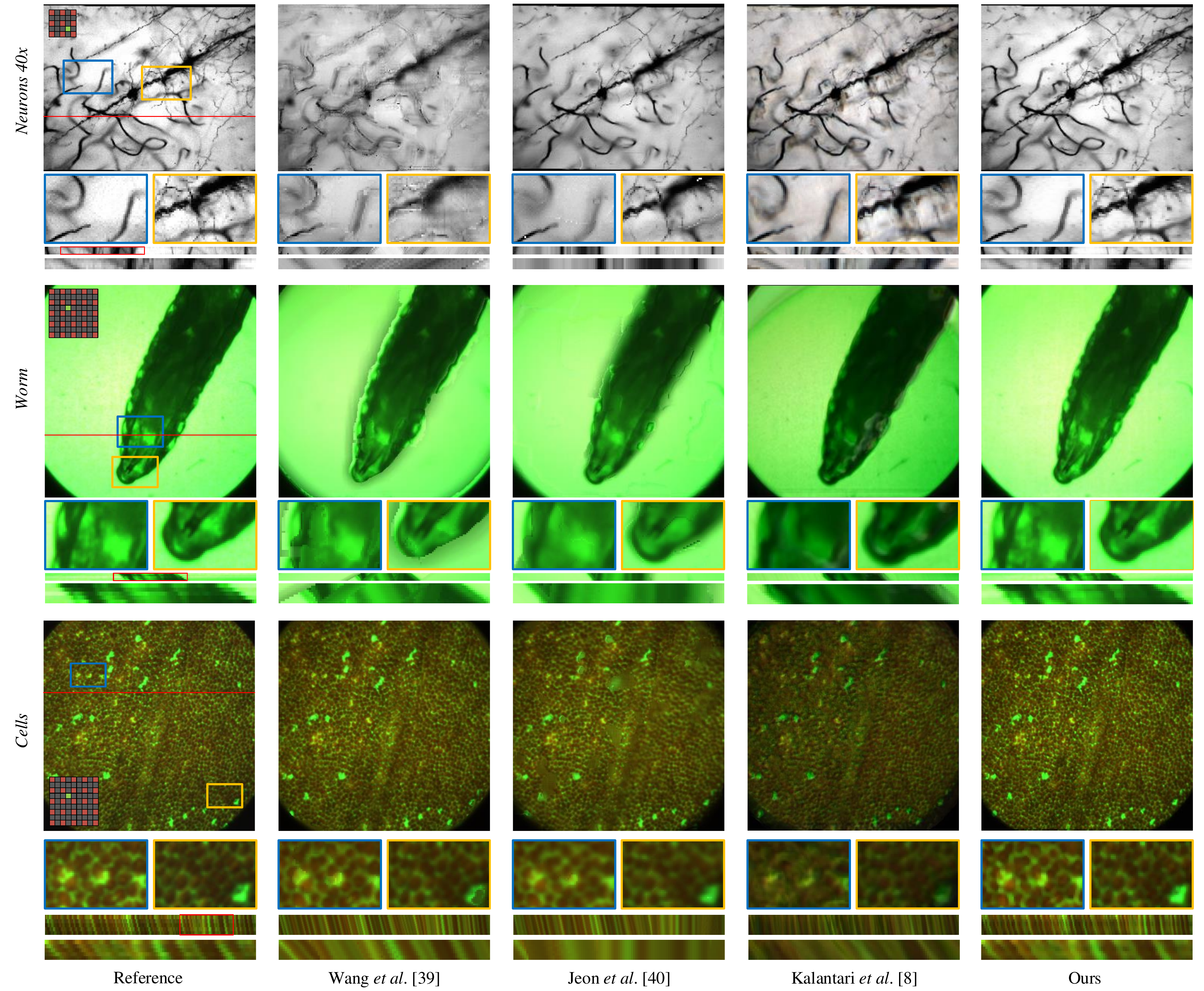}
\end{center}
\vspace{-4mm}
   \caption{Comparison of the proposed approach against other methods on the microscope light field datasets. The results show the ground truth (/ reference images), synthesized results, close-up versions, and the EPIs located at the red line shown in the ground truth view.}
\label{fig:Result3}
\end{figure*}

\begin{table}
\begin{center}
\caption{Quantitative results (PSNR / MS-SSIM) of reconstructed light fields on the real-world scenes. Results from a single CNN is also listed.}
\label{table:Result2}
\vspace{-2mm}
\begin{tabular}{p{2.4cm}p{1.5cm}<{\centering}p{1.5cm}<{\centering}p{1.5cm}<{\centering}}
\hline
& 30 scenes &\textit{Reflective}29&\textit{Occlusion}16 \\
\hline
Wang \textit{et al.} \cite{Occ}& 33.03/0.9766 & 28.97/0.9613 & 25.94/0.9244 \\
Jeon \textit{et al.} \cite{Acc}& 34.42/0.9841 & 40.27/0.9946 & 32.10/0.9830 \\
Kalantari \textit{et al.} \cite{DoubleCNN}& 37.78/0.9912 & 37.70/0.9798 & 32.24/0.9842 \\
\wgcr{Bicubic only} & 34.97/0.9861 & 40.28/0.9952 & 32.97/0.9815 \\
\wgcr{FSRCNN only \cite{FSRCNN}} & 37.23/0.9901 & 43.68/0.9961 & 35.04/0.9848\\
Our CNN only & 37.15/0.9889 & 44.84/0.9962 & 35.89/0.9835\\
Our proposed &\textbf{41.02}/\textbf{0.9968} & \textbf{46.10}/\textbf{0.9981} & \textbf{38.86}/\textbf{0.9970} \\
\hline
\end{tabular}
\end{center}
\end{table}

In this section, we evaluate the proposed framework compared with the approach proposed by Kalantari \textit{et al.} \cite{DoubleCNN} and typical depth-based approaches on several datasets including real-world scenes, microscope light field data and synthetic scenes. For the typical depth-based approaches, we first use current state-of-the-art approaches (Wang \textit{et al.} \cite{Occ}, Jeon \textit{et al.} \cite{Acc}) to estimate the depth; then we warp the input images to the novel view and blend by weighting the warped images \cite{CDSD13}. We also evaluate each step in the framework, including the performance without the ``blur-deblur'' steps (implemented by directly upsampling using learning-based methods) or without the ``restoration'' step (implemented by replacing the learning-based restoration with bicubic interpolation). The quality of the synthesized views is measured by the PSNR and \wgcr{MS-SSIM} \cite{MS-SSIM} against the ground truth image. Comparison results are also given in the submission video \footnote{The first submitted video corresponds to the evaluation part of this paper. Due to the file size limitation, the videos are compressed and shortened. High quality and full videos are available at:  http://www.liuyebin.com/lfepi/LFreconstruction.html}.

\begin{table}
\begin{center}
\caption{Quantitative results (PSNR / MS-SSIM) of reconstructed light fields on the microscope light field datasets.}
\label{table:Result3}
\begin{tabular}{p{2.5cm}p{2.0cm}<{\centering}p{2.0cm}<{\centering}}
\hline
& \textit{Neurons}$\ 20\times$ &\textit{Neurons}$\ 40\times$ \\
\hline
Wang \textit{et al.} \cite{Occ}& 17.45/0.7368 & 13.21/0.7206 \\
Jeon \textit{et al.} \cite{Acc}& 23.02/0.9338 & 23.07/0.9092 \\
Kalantari \textit{et al.} \cite{DoubleCNN}& 20.94/0.9169 & 19.02/0.8847 \\
Our proposed &\textbf{29.34}/\textbf{0.9741} & \textbf{32.47}/\textbf{0.9901}\\
\hline
\end{tabular}
\vspace{-2mm}
\end{center}
\end{table}

\subsection{Real-world scenes}


\begin{table*}
\begin{center}
\caption{Quantitative results (PSNR / MS-SSIM) of the reconstructed light fields on the synthetic scenes of the HCI datasets. The ``restoration'' step can also adopt other learning-based approaches, such as SC \cite{SC} and FSRCNN \cite{FSRCNN}. The bicubic serves as a baseline approach for comparison.}
\label{table:Result1}
\begin{tabular}{p{3.0cm}p{2cm}<{\centering}p{2cm}<{\centering}p{2cm}<{\centering}p{2cm}<{\centering}}
\hline
& \multicolumn{2}{c}{\textit{Buddha}} &\multicolumn{2}{c}{\textit{Mona}}\\
\hline
Angular resolution& $3\times3$ & $5\times5$ & $3\times3$ & $5\times5$\\
\hline
Wang \textit{et al.} \cite{Occ}& 33.41/0.9740 & 44.15/0.9984 & 30.74/0.9720 & 43.69/0.9990\\
Jeon \textit{et al.} \cite{Acc}& 41.19/0.9958 & 44.06/0.9981 & 40.95/0.9968 & 42.67/0.9982\\
Kalantari \textit{et al.} \cite{DoubleCNN}& 34.05/0.9788 & 34.51/0.9813 & 32.53/0.9640 & 32.59/0.9800\\
\wgcr{Ours/bicubic} & 34.71/0.9782 & 39.23/0.9931 & 38.21/0.9870 & 42.66/0.9964\\
Ours/SC \cite{SC}& 41.67/0.9953 & 41.79/0.9962 & 42.39/0.9951 & 44.40/0.9979\\
\wgcr{Ours/FSRCNN} \cite{FSRCNN}& 42.60/0.9960 & 46.27/0.9981 & 43.52/0.9954 & 49.75/0.9992\\
Our proposed &\textbf{43.20}/\textbf{0.9963} & \textbf{46.42}/\textbf{0.9987} & \textbf{44.37}/\textbf{0.9977} & \textbf{51.07}/\textbf{0.9995}\\
\hline
\end{tabular}
\end{center}
\end{table*}

We evaluate the proposed approach using 30 test scenes provided by Kalantari \textit{et al.} \cite{DoubleCNN} that were captured with a Lytro Illum camera (``30 scenes'' for short) as well as two representative scenes, \textit{Reflective} 29 and \textit{Occlusion} 18, from the Stanford Lytro Light Field Achieve \cite{StanfordLytro}. We use $3\times3$ views to reconstruct $7\times7$ light fields. \wgcr{The running times for each step are as follows: the ``blur'' step takes $2.42\times10^{-4}$ seconds per view, the ``restoration'' step takes $10.10$ seconds per view and the ``deblur'' step takes $4.46$ seconds per view. The hardware configuration is the same as that described in Section 4.3.2, but GPU is not involved in each step.}

Table \ref{table:Result2} lists the numerical results on the real-world datasets. The PSNR values are averaged over the 30 scenes. The CNNs in the approach by Kalantari \textit{et al.} \cite{DoubleCNN} are designed to minimize the error between the synthesized views and the ground truth views \footnote{\wgcr{To follow the configuration of their four corner images input, we divide the input light field into several blocks accordingly, e.g., a 3 by 3 input light field is divided into four light fields, where each contains 2 by 2 sub-aperture images for the angular super-resolution.}}. Therefore, they achieve better performance than other depth-based methods when applied to those common scenes. However, their networks were specifically trained for Lambertian regions; thus, they tend to fail on the reflective surface in the \textit{Reflective} 29 case. Among these real-world scenes, our proposed framework is significantly better than other approaches. \wgcr{We also compare the results produced by using single CNNs without the ``blur-deblur'' framework (please see Section \ref{Sec:FA} for the detailed discussion), including our network (denoted as ``Our CNN only'') and FSRCNN \cite{FSRCNN} (been fine-tuned on EPIs, denoted as ``FSRCNN only''). The quantitative results show that a single CNN produces lower quality light fields than those using the complete framework.}

Figure \ref{fig:Result2} depicts some of the results such as the \textit{Leaves} from the 30 scenes as well as \textit{Reflective} 29 and \textit{Occlusion} 16 scenes in the Stanford Lytro Light Field Archive. The \textit{Leaves} case includes some leaves with complex structures in front of a street. The case is challenging due to the overexposure of the sky and the occlusion around the leaves shown in the blue box. The results by Wang \textit{et al.} \cite{Occ} and Jeon \textit{et al.} \cite{Acc} show blurring artifacts around the leaves, and the result by Kalantari \textit{et al.} \cite{DoubleCNN} contains ghosting artifacts. The \textit{Reflective} 29 case is a challenge scene because of the reflective surfaces of the pot and the kettle. The result by Wang \textit{et al.} \cite{Occ} shows blurring artifacts around the pot and the kettle. The approaches by Jeon \textit{et al.} \cite{Acc} and Kalantari \textit{et al.} \cite{DoubleCNN} produce better results, but the reconstructed light fields show discontinuities in terms of the EPIs. The \textit{Occlusion} 16 case contains complicated occlusions that are challenging for view synthesis; consequently, their results are quite blurry around occluded regions such as the branches and leaves. As demonstrated in the error maps and the close-up images of the results, the proposed approach achieves a high performance in terms of the visual coherency of both the synthesized views and the EPIs.

\subsection{Microscope light field dataset}

In this subsection, the Stanford Light Field microscope datasets \cite{microLF} and the camera-array-based light field microscope datasets provided by Lin \textit{et al.} \cite{microLFArray} are tested. These datasets include challenging light fields such as those containing complicated occlusion relations and translucency. The numerical results are tabulated in Table \ref{table:Result3}, and the reconstructed views are shown in Figure \ref{fig:Result3}. We reconstruct $7\times7$ light fields using $3\times3$ views in the \textit{Neurons} $40\times$ case, and $5\times5$ light fields using $3\times3$ views in the \textit{Neurons} $40\times$ case. For the \textit{Worm} and \textit{Cells} cases, $5\times5$ views are used to produce $9\times9$ light fields\footnote{The quantitative evaluation is not performed on the \textit{Worm} and \textit{Cells} cases because all the ground truth views are used as input.}.

The \textit{Neurons} $40\times$ case shows a Golgi-stained slice of rat brain, which contains complicated occlusion relations. The result by Wang \textit{et al.} \cite{Occ} is quite blurry due to the errors in the estimated depth. Although the result by Jeon \textit{et al.} \cite{Acc} has a higher PSNR value, it fails to estimate the depth of the scene, which is visible in the EPI. The result produced by Kalantari \textit{et al.} \cite{DoubleCNN} has a higher quality in terms of the visual coherency. However, the result contains blurring and tearing artifacts in the occluded regions. The \textit{Worm} and \textit{Cells} cases are more simply structured but contain transparent objects such as the head of the worm. The depth-based approaches are unable to estimate accurate depth maps in the translucent regions, which results in tearing and ghosting artifacts.  Among these challenging cases, our approach produces plausible results in both the occluded and translucent regions.

\subsection{Synthetic scenes}

We use the synthetic light field data from the HCI datasets \cite{HCI} in which the spatial resolution is the same as the original inputs ($768\times768$). The angular resolution of the output light field is set to $9\times9$ for comparison with the ground truth images, although we are able to produce light fields of denser views. We use input light fields with different degrees of sparsity ($3\times3$ and $5\times5$) to evaluate the performance of the proposed framework for different upsampling scale factors. The results show that our framework is more competent for inputs with different degrees of sparsity. Table \ref{table:Result1} shows a quantitative evaluation of the proposed approach on the synthetic dataset compared with other methods. The approach by Kalantari \textit{et al.} \cite{DoubleCNN} produces a lower quality than the other depth-based approaches because their CNNs are specifically trained on real-world scenes. The proposed approach achieves the highest PSNR values compared to the depth-based methods. \wgcr{In addition, we replace our residual network with bicubic interpolation as a baseline approach to demonstrate the importance of the learning-based "restoration" step. The results show that the reconstruction benefits from the learning-based ``restoration'' step (such as SC \cite{SC}, FSRCNN \cite{FSRCNN} and the proposed network), especially when using a sparser input.}


\section{Extended Applications}

We implement three different applications based on the proposed ``blur-restoration-deblur'' framework: depth enhancement using reconstructed high-angular-resolution light fields, \wgcb{interpolation for unstructured input and depth-assisted rendering}.

\subsection{Application for depth enhancement}

\begin{figure}
\begin{center}
\includegraphics[width=0.95\linewidth]{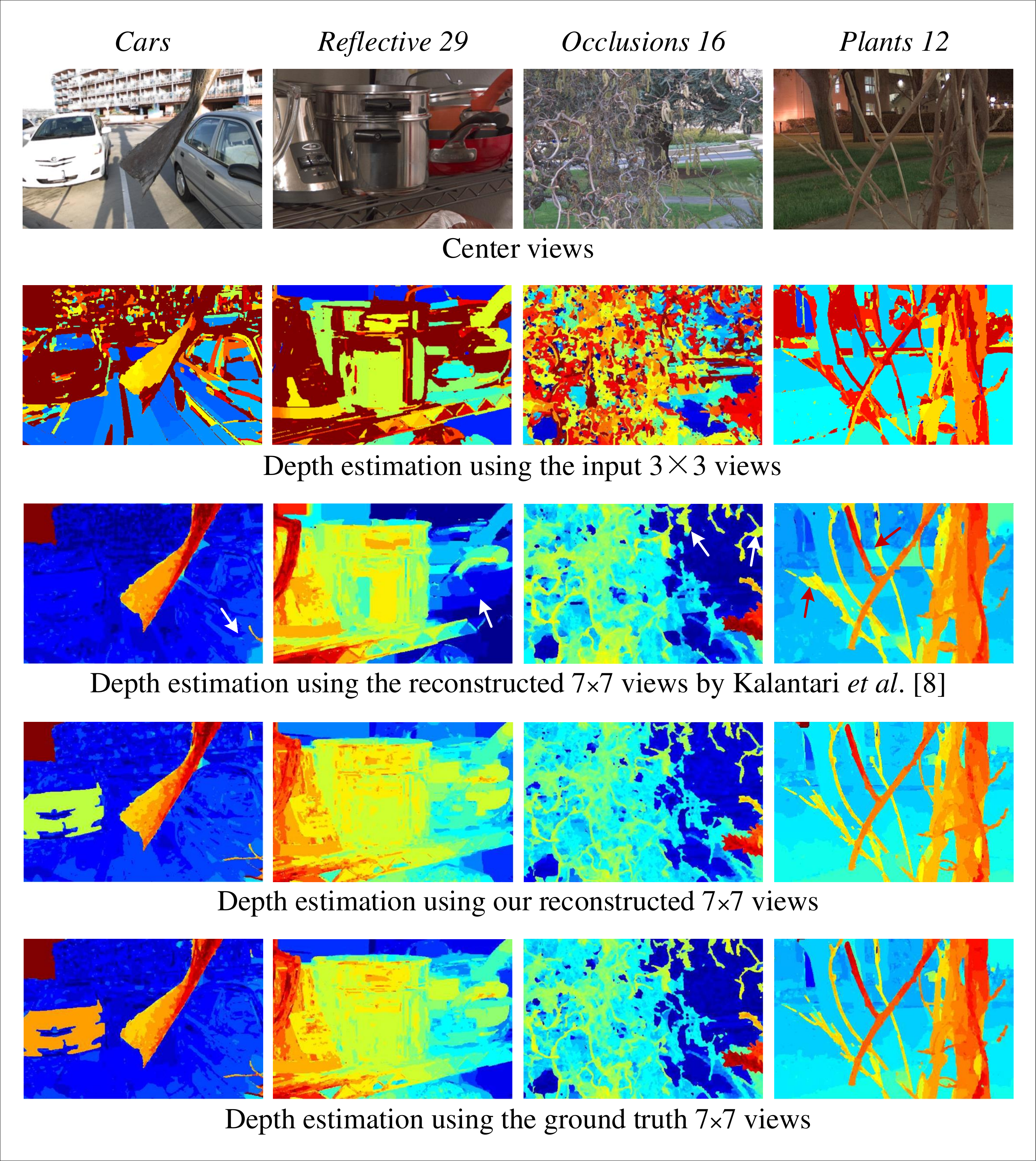}
\end{center}
\vspace{-4mm}
   \caption{Depth estimation results using the reconstructed light fields. The arrows in the third row indicate the depth errors caused by the artifacts of the reconstructed light fields.}
\label{fig:App1}
\end{figure}

\begin{table}
\begin{center}
\caption{RMSE values of the estimated depth using the approach by Wang \textit{et al.} \cite{Occ} on HCI datasets.}
\label{table:App1}
\begin{tabular}{p{2.4cm}p{1.4cm}<{\centering}p{1.4cm}<{\centering}p{1.4cm}<{\centering}}
\hline
& \textit{Buddha} &\textit{Mona} &\textit{Horses}\\
\hline
Input $3\times3$ views& 0.2926 & 0.2541 & 0.3757\\
Kalantari \textit{et al.} \cite{DoubleCNN}& 0.1576 & 0.0829 & 0.1212\\
Ours & 0.0401 & 0.0517 & 0.0426 \\
GT light fields &0.0393 & 0.0529 & 0.0383\\
\hline
\end{tabular}
\end{center}
\vspace{-2mm}
\end{table}

This application demonstrates that the reconstructed high-angular-resolution light field can be applied to enhance the depth estimation. Both synthetic scenes and real-world scenes are used for the demonstration. We use the approach by Wang \textit{et al.} \cite{Occ} to estimate the depth of the scenes.

For synthetic scenes, Table \ref{table:App1} gives the RMSE values of the depth estimation results from using the $3\times3$ light fields, the reconstructed $9\times9$ light fields produced by Kalantari \textit{et al.} \cite{DoubleCNN}, our reconstructed $9\times9$ light fields and the ground truth $9\times9$ light fields on the HCI datasets \cite{HCI}. Figure \ref{fig:App1} shows the depth estimation results on some real-world scenes, including the \textit{Cars}, \textit{Reflective} 29, \textit{Occlusion} 16, and \textit{Flowers and plants} 12 cases. The results show that our reconstructed light fields are able to produce more accurate depth maps that better preserve edge information than the those produced by Kalantari \textit{et al.} \cite{DoubleCNN}, e.g., the reflective surface of the red pan in the \textit{Reflective} 29 case and the branches in front of the left car in the \textit{Cars} case. Moreover, the enhanced depth maps are highly similar to the maps produced using the ground truth light fields, which once again demonstrates that the proposed framework is able to preserve the structure of the reconstructed light fields better than anther methods.

\subsection{\wgcb{Interpolation for unstructured light field}}

\begin{figure}
\begin{center}
\includegraphics[width=1\linewidth]{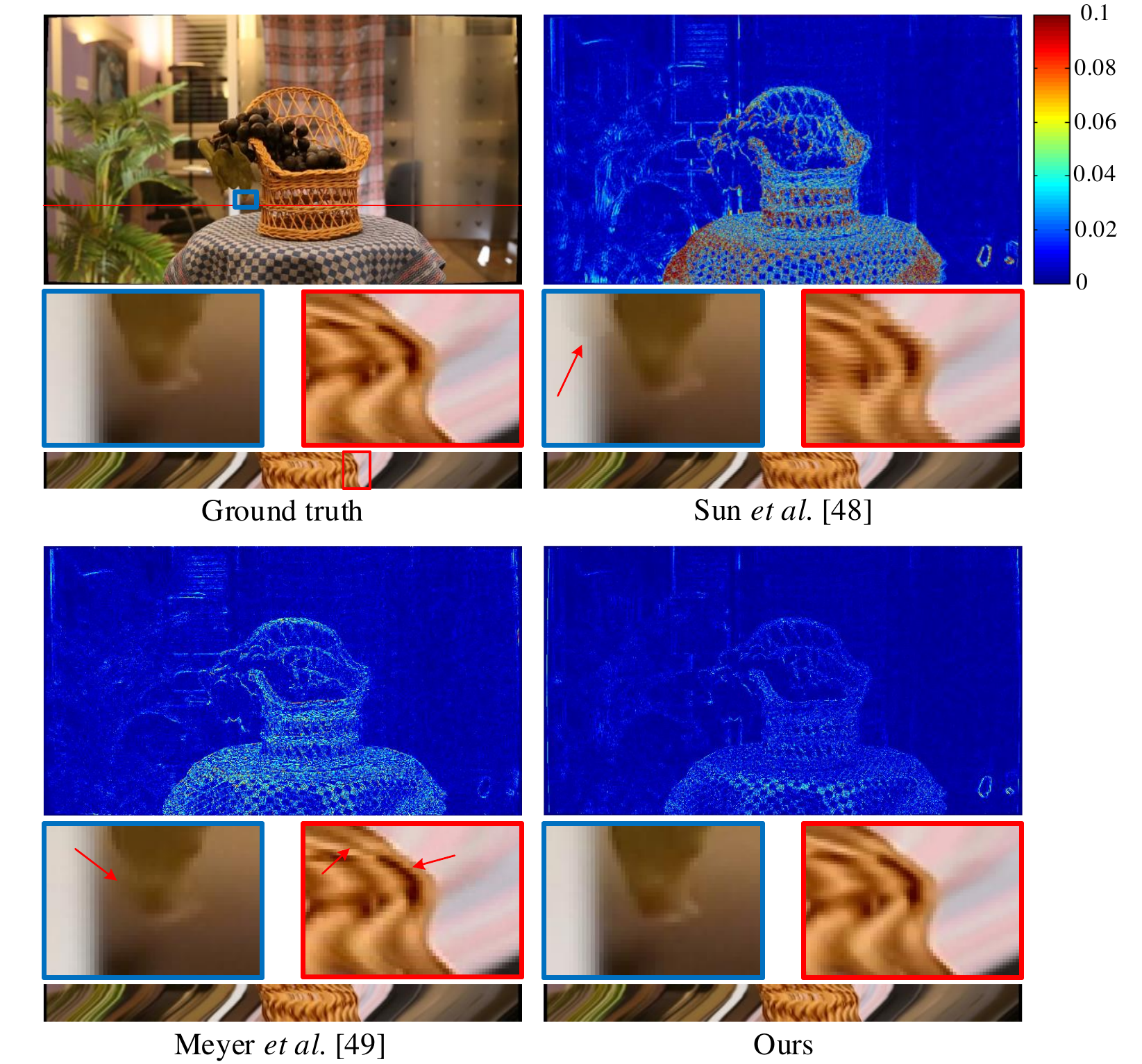}
\end{center}
\vspace{-4mm}
   \caption{\wgcb{Comparison of unstructured light field super-resolution against the approaches by Sun \textit{et al.} \cite{classicNL} and Meyer \textit{et al.} \cite{Meyer15}. \textit{Basket} dataset courtesy of Y\"{u}cer \textit{et al.} \cite{Kaan16}.}}
\label{fig:ULF}
\end{figure}

\wgcb{The proposed approach is a depth-free framework that restores the angular details based on CNN without the need for geometry calibration or depth estimation. Therefore, our framework is potentially capable of handling unstructured input such as an unstructured light field \wgcr{(3D light field)}. The interpolated unstructured light field can be further applied to improve the accuracy of the reconstructed 3D model \cite{Kaan16}. In this section, we compare the proposed framework against the optical flow-based approach by Sun \textit{et al.} \cite{classicNL} and the phase-based interpolation approach proposed by Meyer \textit{et al.} \cite{Meyer15}. To obtain the interpolated views based on the computed optical flow maps, the DIBR technique proposed by Riechert \textit{et al.} \cite{riechert2012fully} is applied.}


\wgcb{Unlike common light fields that are obtained by carefully calibrated camera(s), an unstructured light field is captured, e.g., by a hand-held commodity camera \cite{Davis2012} or unsynchronized camera arrays \cite{wang2007space}. Therefore, ordinary angular super-resolution methods that synthesize novel views based on depth estimation usually fail to yield reasonable results. On the contrary, the proposed framework implicitly reconstructs a light field by restoring the angular details on the EPI, and thus, it has the capacity to super-resolve unstructured light fields.}

\wgcb{We demonstrate this application using the \textit{Basket} case from the dataset provided by Y\"{u}cer \textit{et al.} \cite{Kaan16} (see Figure \ref{fig:ULF}). The original light field contains 49 views, of which 25 views are used as input for the super-resolution. The averaged PSNR values are 33.82 for the approach by Sun \textit{et al.} \cite{classicNL}, 40.43 for the approach by Meyer \textit{et al.} \cite{Meyer15} and 41.60 for the proposed framework. Figure \ref{fig:ULF} shows interpolation results of the 20$^{\text{th}}$ frame. The results by Sun \textit{et al.} and Meyer \textit{et al.} \cite{Meyer15} introduce structure discontinuity (see the close-up version of the EPI) because only two views are used for each interpolation.}

\begin{figure}
\begin{center}
\includegraphics[width=1\linewidth]{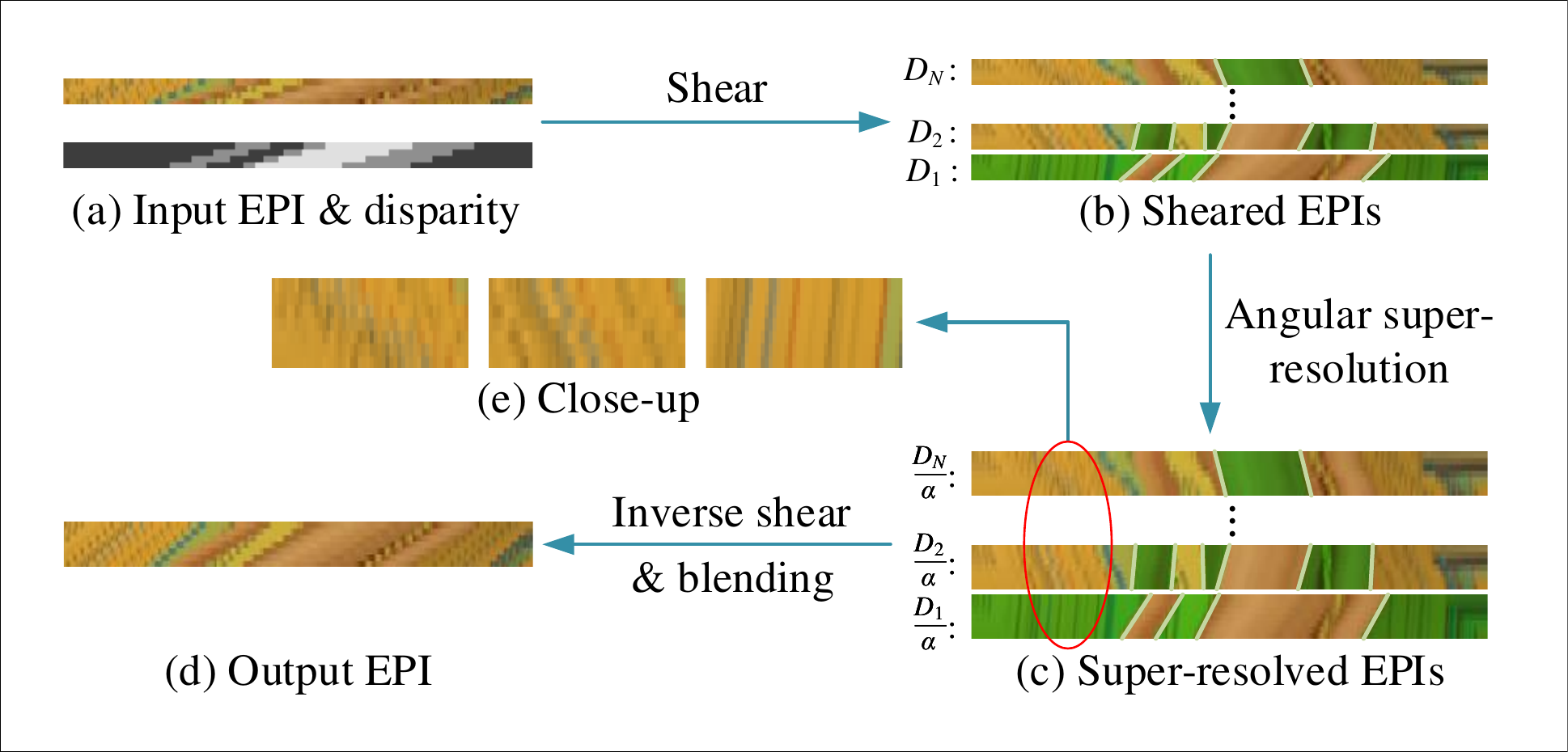}
\end{center}
\vspace{-4mm}
   \caption{\wgcb{Pipeline of the proposed depth-assisted rendering approach using the ``blur-restoration-deblur'' framework. (a) The input EPI and its discretized disparity; (b) The input EPI is sheared by each shear value, thereby constructing a set of sheared EPIs. The regions marked in green masks are sheared using their corresponding disparity, but other regions are not; (c) The super-resolved EPIs using the proposed ``blur-restoration-deblur'' framework; (d) The final high-angular-resolution EPI is obtained using the inverse shear and blending operation; (e) Close-up of the super-resolved EPIs in (c).}}
\label{fig:DARframework}
\end{figure}

\subsection{\wgcb{Depth-assisted rendering}}

\wgcb{We further extend the proposed approach for novel view rendering using multi-view input with an estimated depth (disparity) map. By exploiting the disparity information, our proposed method can achieve high-quality view rendering for the cases with larger disparities (up to 40 pixels). Unlike existing depth-image-based rendering (DIBR) techniques, which are extremely sensitive to the disparity quality, our method only requires a raw disparity map to produce high-quality rendering results. The proposed novel view rendering method also inherits the ability to generate plausible results in occluded regions, and non-Lambertian surfaces from the proposed learning-based framework.}

\wgcb{As described, the main challenge in reconstructing an EPI is the information asymmetry, which needed to be balanced by the ``blur'' step. However, for an EPI with large disparity, the size of the blur kernel should be sufficiently large to remove the high-frequency leakage, which will also influence the image quality. Alternatively, we shear it to an appropriate disparity range with the assistance of its corresponding disparity map such that the proposed ``blur-restoration-deblur'' framework can be applied.}

\wgcb{\textbf{Implementation.} Consider an EPI $E_L$ (where $L$ denotes low angular resolution) and its discretized disparity $D$ (see Figure \ref{fig:DARframework}(a)). The collection of shear values is equivalent to the collection of discretized disparity values in $D$, denoted as $\{D_1,D_2,...,D_N\}$ ($N$ is the number of shear values). We first shear the EPI by each shear value, thereby constructing a set of sheared EPIs $\{S(E_L^{D_1}),S(E_L^{D_2}),...,S(E_L^{D_N})\}$ (see Figure \ref{fig:DARframework}(b)), where $S$ denotes the shear operation and $N$ is the number of the discretized disparity values. Then the proposed ``blur-restoration-deblur'' framework is applied to super-resolve the sheared EPIs in the angular dimension (see Figure \ref{fig:DARframework}(c)), generating a set of high-angular-resolution EPIs $\{S(E_H^{D_1}),S(E_H^{D_2}),...,S(E_H^{D_N})\}$. For regions sheared by their corresponding disparity (indicated as the regions marked in green masks in Figure \ref{fig:DARframework}(b) and (c)), we can obtain super-resolved results with the desired quality; however, other regions are sheared by unbefitting shear values, therein causing aliasing effects in the corresponding regions of the super-resolved EPIs. Figure \ref{fig:DARframework}(e) shows a close-up of one of those super-resolved regions. To combine all the best reconstructed regions into the final EPI $E_H$, we apply the following steps: the super-resolved EPIs $\{S(E_H^{D_1}),S(E_H^{D_2}),...,S(E_H^{D_N})\}$ are first inversely sheared by the corresponding shear values $\{\frac{D_1}{\alpha},\frac{D_2}{\alpha},...,\frac{D_N}{\alpha}\}$, where $\alpha$ is the super-resolution factor; the processed EPIs, denoted as $\{E_H^{D_1},E_H^{D_2},...,E_H^{D_N}\}$, are then blended using the following equation:
\begin{equation} \label{Eq:blending}
E_H=\sum_{i=1}^{N}E_H^{D_i} W_i,
\end{equation}
where the weight $W_i$ is equal to $1$ for regions sheared by the corresponding disparity and $0$ for other regions. Figure \ref{fig:DARframework}(d) shows the final high-angular-resolution EPI $E_H$.}

\begin{figure*}
\begin{center}
\includegraphics[width=1\linewidth]{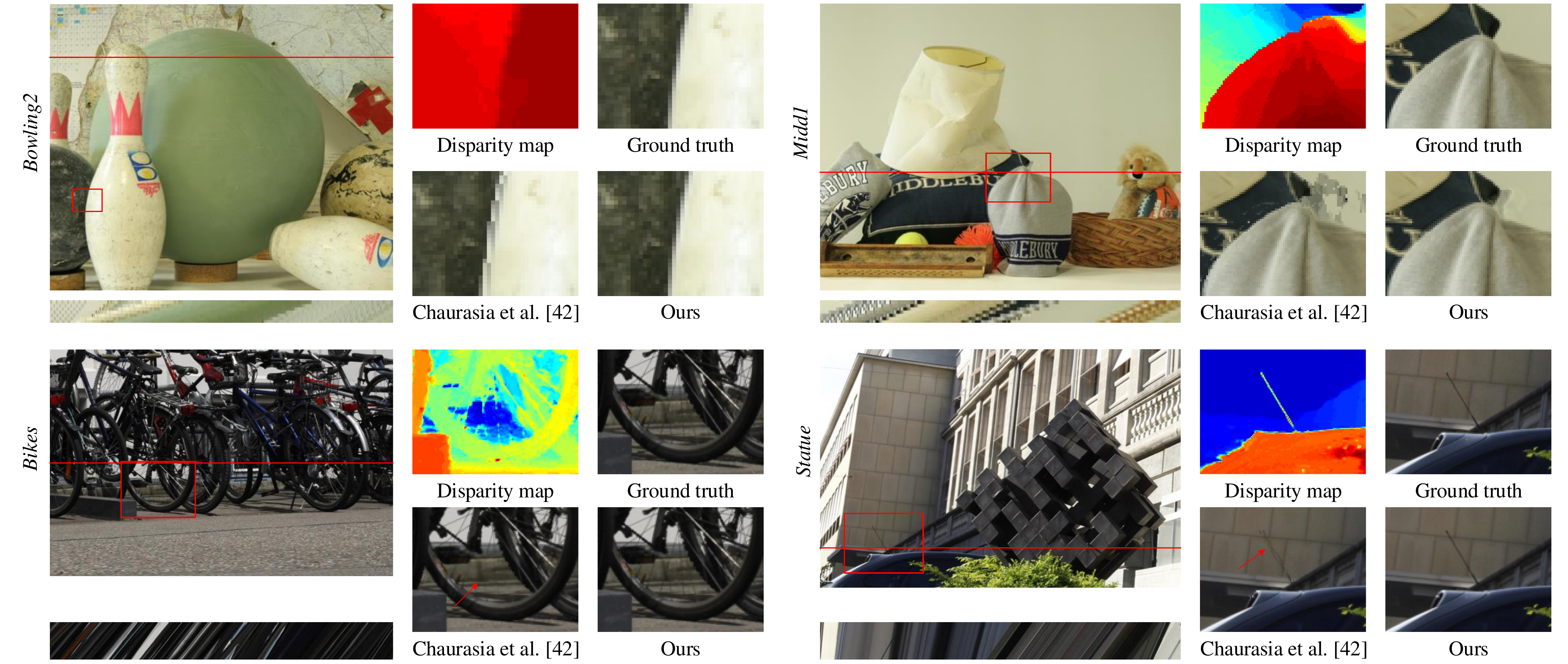}
\end{center}
\vspace{-4mm}
   \caption{\wgcb{Depth-assisted rendering results. The left images show the rendered views (the 4$^{\text{th}}$ view for \textit{Bowling2} and \textit{Midd1} and the 39$^{\text{th}}$ view for \textit{Bikes} and \textit{Statue}) and the corresponding EPIs using the proposed approach. The close-up versions of reference disparity maps, ground truth views and rendered views by Chaurasia \textit{et al.} \cite{CDSD13} and the proposed approach are shown on the right.}}
\label{fig:DARresult1}
\end{figure*}

\begin{table*}
\begin{center}
\caption{\wgcb{Quantitative results of the disparity map range (pixel) and error (RMSE) and rendered views (PSNR / MS-SSIM). The image rendering results are averaged on the novel views.}}
\label{table:DARresult}
\begin{tabular}{lccccccc}
\hline
Dataset& \textit{Baby1} &\textit{Bowling2} &\textit{Moebius} & \textit{Midd1} &\textit{Monopoly} &\textit{Bikes} &\textit{Statue}\\
\hline
\multicolumn{8}{c}{Disparity map}\\
\hline
Disparity range (pixel) & 8-23 & 11-34 & 11-35 & 9-33 & 2-27 & 0-40 & 0-24\\
Disparity error (RMSE) & 0.0369 & 0.0401 & 0.0473 & 0.1682 & 0.1849 & - & -\\
\hline
\multicolumn{8}{c}{Rendered views (PSNR / MS-SSIM)}\\
\hline
Riechert \textit{et al.} \cite{riechert2012fully} & 36.56/0.9880 & 34.19/0.9869 & 33.06/0.9860 & 31.18/0.9757 & 30.99/0.9770 & 25.65/0.9047 & 35.57/0.9958\\
Chaurasia \textit{et al.} \cite{CDSD13} & 37.90/0.9914 & 35.44/0.9907 & 34.35/0.9910 & 33.67/0.9808 & 32.01/0.9881 & 30.90/0.9612 & 35.04/0.9965\\
Ours &\textbf{40.92}/\textbf{0.9949} & \textbf{38.22}/\textbf{0.9937} & \textbf{38.41}/\textbf{0.9953} & \textbf{38.46}/\textbf{0.9969} & \textbf{40.04}/\textbf{0.9938} & \textbf{35.43}/\textbf{0.9890} & \textbf{36.44}/\textbf{0.9968}\\
\hline
\end{tabular}
\end{center}
\end{table*}

\wgcb{\textbf{Results.} We evaluate the proposed application for depth-assisted rendering on the Middleburry stereo datasets \cite{scharstein2002taxonomy} and outdoor light fields \cite{kim2013scene}. For the Middleburry stereo datasets, we employ the CostFilter \cite{CostFilter} for the disparity estimation. In addition, the disparity maps for the outdoor light fields are taken from \cite{kim2013scene}. The DIBR techniques proposed by Riechert \textit{et al.} \cite{riechert2012fully} and Chaurasia \textit{et al.} \cite{CDSD13} are used as the baseline methods. Each scene in the Middleburry stereo datasets contains 7 views, 4 of which are used as input and the remainder used for comparison. And for the outdoor light fields, we use 49 views for each scene, 13 of which are used to reconstruct light fields with original angular resolution.}

\wgcb{Figure \ref{fig:DARresult1} shows a visual comparison of the rendered novel views against the ground truth, and Table \ref{table:DARresult} offers the relevant numerical results in terms of disparity quality (RMSE) and novel view quality (PSNR and MS-SSIM). \textit{Baby1}, \textit{Bowling2}, \textit{Moebius}, \textit{Midd1} and \textit{Monopoly} are from the Middleburry stereo datasets \cite{scharstein2002taxonomy}. \textit{Bikes} and \textit{Statue} are taken from Kim \textit{et al.} \cite{kim2013scene}. As we can see from the figure, the approach by Chaurasia \textit{et al.} \cite{CDSD13} contains ghosting artifacts in occlusion boundaries (see the close-up in \textit{Bowling2} and \textit{Midd1} in Figure \ref{fig:DARresult1}) and blur artifacts around small structures (such as the spokes in \textit{Bikes} and the antenna in \textit{Statue} in Figure \ref{fig:DARresult1}). In addition, due to the extensive textureless regions in the last two cases (\textit{Midd1} and \textit{Monopoly}), the stereo matching approach being employed has failed to produce plausible disparity maps, which greatly affects the results by DIBR methods, especially the one by Riechert \textit{et al.} \cite{riechert2012fully}. However, due to the robustness to disparity noise, the proposed approach is able to render high-quality views in both cases. Table \ref{table:DARresult} also lists the disparity range of each dataset, showing the good disparity handling ability of the proposed rendering approach.}

\section{Discussion and Conclusion}

\wgcb{In the following, we will provide an in-depth analysis on the merits of the learning-based reconstruction using EPIs. Due to the high-dimensional data in a light field, the acquisition process suffers from a resolution trade-off. Relatively speaking, it is a better choice to sacrifice the angle resolution and obtain more spatial information, which is also better for the reconstruction of the light field. However, essentially, the reconstruction is an ill-posed problem to restore the plenoptic function from a limited collection of samples, which is typically solved by applying strong prior information. Therefore, the deep neural networks learned from massive training data may represent a good solution.}

\wgcb{Deep neural networks work well for data sharing similar properties with the training data, but they are heavily data dependent. In sub-aperture image space, the data structures, namely, the appearances of the scenes, could be very different (e.g., the real-world scenes in Figure \ref{fig:Result2} and the microscopy scenes in Figure \ref{fig:Result3}, where even different microscopy data shows very different appearance). It is difficult to generalize from such a variety of scenes. Thus, training a network using sub-aperture images maybe not the best solution. Fortunately, in the EPI space, even the light fields of different scenes share strikingly similar EPI structures because the angular dimension in the EPI records the information of the same pixel in the 3D space with changing viewpoint. Therefore, this paper have taken advantage of the EPI property to achieve high performance light field angular super-resolution. The light field reconstruction on the EPI provides a higher capacity in the data structure term. For the above reasons, the proposed learning-based framework for light field reconstruction on the EPI achieves a better performance on various scenes.}

In this paper, we have presented a novel ``blur-restoration-deblur'' framework for light field reconstruction on EPI and its extended applications. To avoid the aliasing effects caused by the information asymmetry, the spatial low-frequency components of the EPI are extracted via an EPI blur operation and used as input to the network to restore the angular details. The non-blind deblur operation is used to recover the spatial details that are suppressed by the EPI blur operation. We evaluate the proposed framework on synthetic scenes, real-world scenes, and some challenging microscope light field datasets. The experimental results demonstrate that the proposed framework outperforms state-of-the-art approaches in occluded and transparent regions and on non-Lambertian surfaces. \wgcb{The results (Table \ref{table:Result2} and Table \ref{table:Result1}) also show that both ``blur-deblur'' steps and learning-based ``restoration'' are important to the proposed framework.} We further show extended applications, including depth enhancement using reconstructed high-angular-resolution light fields, interpolation for unstructured input and depth-assisted rendering.

\begin{figure}
\begin{center}
\includegraphics[width=1\linewidth]{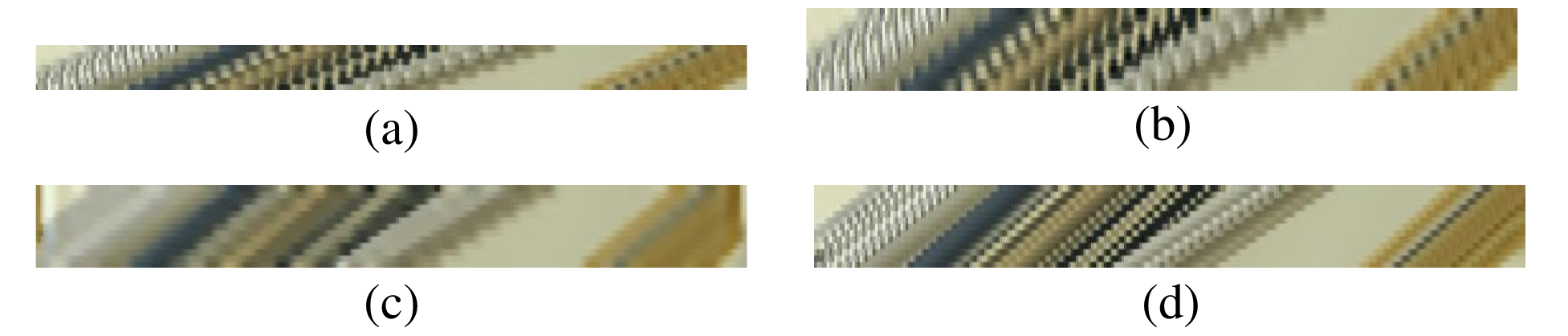}
\end{center}
\vspace{-4mm}
   \caption{\wgcr{Limitation of reconstructing an input EPI with large disparity. (a) An input EPI extracted from multi-view stereo data \cite{scharstein2002taxonomy}, \textit{Midd1}; (b) Result produced by Straightforward CNN-based angular super-resolution, which appears severe angular aliasing effects; (c) Result produced by the proposed framework. The details are difficult to recover, because a blur kernel of large size has to be used for anti-aliasing; (d) Result produced by the proposed depth-assisted rendering approach.}}
\label{fig:Limitation}
\end{figure}

In the following, the limitations of the proposed framework that should be overcome in the future work are concluded. \wgcr{The framework adopts at least three views in each angular dimension for the initial interpolation, and extrapolation cannot be handled in the current implementation}. We use EPI blur to extract the spatial low-frequency components of the EPI, where the kernel size is determined by the largest disparity between the input neighboring views. \wgcr{The non-blind deblur is unable to recover high quality EPIs when the kernel size is too large (as shown in Figure \ref{fig:Limitation}(c))}, and the maximum disparity we can handle when using the proposed ``blur-restoration-deblur'' framework is 5 pixels. \wgcb{However, in our extended application, we exploit depth information to handle large disparity data such as multi-view stereo data (as shown in Figure \ref{fig:Limitation}(d)).} \wgcr{For reconstructing an unstructured light field, we assume that the vertical disparities (perpendicular to the main direction of the parallax) between the neighboring views are less than one pixel, because the ``blur'' and ``deblur'' steps are designed to address the disparities in only one direction.}

\wgcb{For data degradation in terms of spatial aliasing, the framework will fail to provide reasonable results when the frequency of the texture area is higher than the spatial sampling rate because the framework is not designed for such input. However, the proposed framework presents a denoising effect for data degradation in terms of noise, due to the CNN's inherent noise suppression capability. The denoising effect is clearly reflected in the \textit{Neurons} $20\times$ (Figure \ref{fig:Teaser}) and \textit{Neurons} $40\times$ (Figure \ref{fig:Result3}) cases, where the output light field is substantially smoother than the input light field (see the close-up version).}

\ifCLASSOPTIONcompsoc
  \section*{Acknowledgments}

\else
  \section*{Acknowledgment}
\fi
This work was supported by the National key foundation for exploring scientific instrument No.2013YQ140517, the NSF of China emergency management project No. 61550002, the NSF of China grant No.61522111, No.61531014, No.61673095, No.61722209 and No.61331015.

\ifCLASSOPTIONcaptionsoff
  \newpage
\fi



\bibliographystyle{IEEEtran}
\bibliography{IEEEabrv}
%
%
%

%




\end{document}